\documentclass[aps,pra,longbibliography,superscriptaddress,twocolumn,twocolumn,10pt]{revtex4-1}
\usepackage{amsmath}
\usepackage{latexsym}
\usepackage{graphicx}
\usepackage{amsfonts}
\usepackage{braket}
\usepackage{hyperref}
\usepackage{natbib}
\usepackage{amssymb}
\usepackage{nicefrac}
\usepackage{fancyhdr}
\usepackage[utf8]{inputenc}
\usepackage{dsfont}
\usepackage{colortbl}
\usepackage{xcolor}
\usepackage{subfigure}
\usepackage{psfrag}
\usepackage{tikz}
\usepackage{paralist}
\usepackage{nicefrac}
\usepackage{ulem}
\usepackage{placeins}
\usepackage{mathtools}
\usepackage{bbm}
\usepackage{color}

\usetikzlibrary{arrows,shapes,decorations.pathmorphing}

\makeindex

\begin{document}

\title{Nonpairwise interactions induced by virtual transitions in four coupled artificial atoms}

\author{M. Sch\"ondorf}
\author{F. K. Wilhelm}
\affiliation{Theoretical Physics, Saarland University, 66123
Saarbr{\"u}cken, Germany}

\begin{abstract}
Various protocols implementing a quantum computer are being pursued, one of which is the adiabatic quantum computer. Natural interactions in electromagnetic environments are only two-local, but the construction or simulation of higher order couplers is indispensable for a universal adiabatic quantum computer using conventional flux qubits (no non-stoquastic interactions). Here we show that in a specific flux qubit coupler design without ancilla qubits, four body stoquastic interactions are induced by virtual coupler excitations. For specific parameter regimes they are the leading effect and can be tuned up to the GHz range.
\end{abstract}

\maketitle

\section{Introduction}
\label{sec:1}
Quantum computers  have the potential to lead to an exponentially reduced computation time compared to classical computers for certain problems. One promising candidate for the realization of such a device is an adiabatic quantum computer (AQC), where the computation proceeds from an initial Hamiltonian whose ground state is easy to prepare to the ground state of a final Hamiltonian which encodes the solution of the computational problem, by avoiding excitations \cite{farhi2000quantum,farhi2001quantum,van2001powerful,RevModPhys.90.015002}. It is now known that an AQC represents a universal quantum computer \cite{aharonov2004}. Still, implementing an AQC with verifiable speedup is a difficult task. A big step is to overcome the locality of natural interactions. $k$ local interactions with $k>2$ are suitable for the effective implementation of various optimization algorithms \cite{bravyi2008quantum}. Furthermore, since conventional qubit designs are not feasible to implement non-stoquastic interactions \cite{vinci2017non, kerman2018superconducting}, a logical way to establish universality in AQCs is realizing interactions with $k>2$ \cite{aharonov2008adiabatic}. There are embedding schemes that simulate this type of coupling requiring large overhead, such that it would be desirable to implement them as natively as possible. Furthermore higher order local interactions are interesting from a fundamental physics point of view, since the only known and proven interaction between more than two particles is found in Efimov states \cite{efimov1971weakly,kraemer2006evidence}.

\begin{figure*}[t!]
\includegraphics[width=.99\textwidth]{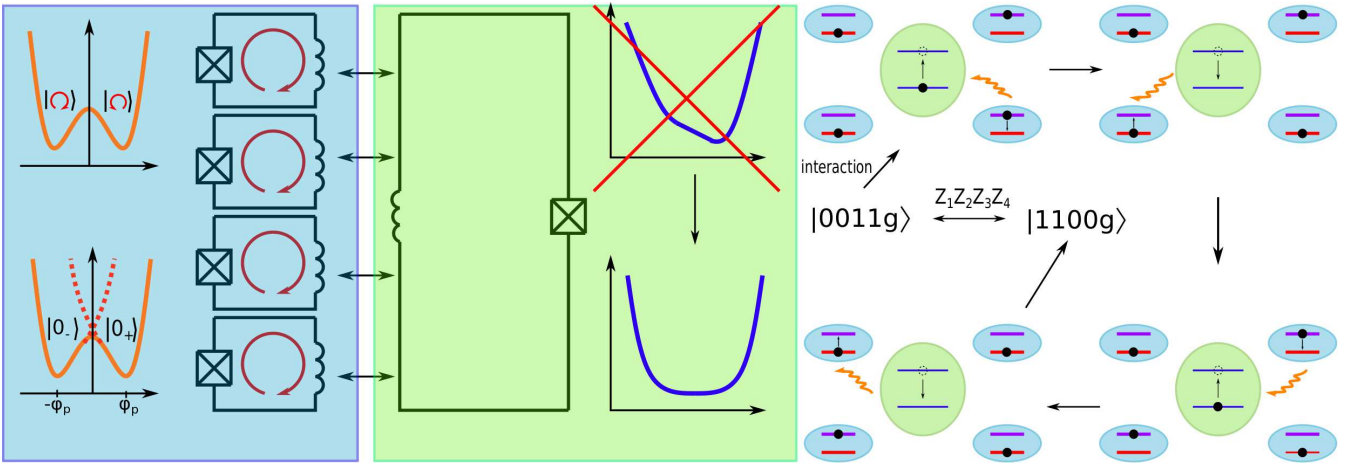}
\caption{\textit{Left: Setup of the coupler architecture. Four flux qubits are inductively coupled to an additional flux qubit with higher energy via mutual inductances $M_j$. We see the qubits double well potential which can be approximated by two shifted harmonic potentials with eigenstates $\ket{0_-}$ and $\ket{0_+}$ corresponding to the persistent current states $\ket{\circlearrowright}$ and $\ket{\circlearrowleft}$ of the qubits. We choose a symmetric coupler potential, to reduce $\hat Z$ corrections arising from $\hat H_{\rm int}$ by biasing the coupler at the flux degeneracy point, as shown on the right of the coupler.} Right: Visualization of four local interaction induced by virtual coupler transitions. A virtual excitation and annihilation of the coupler leads to an energy transfer between two qubits, hence two such processes can induce effective four local interactions.}
\label{setup}
\end{figure*}
There are many proposals to realize higher local interactions, using quantum embedding or ancilla qubits \cite{jordan2008perturbative,chancellor2017circuit,leib2016transmon}, which all create a huge overhead. In this paper we propose a specific coupler architecture using flux qubits \cite{mooij1999josephson} and prove the existence of non-negligible antiferromagnetic four local interactions. 

In Sec. \ref{sec:2} we present the system of interest and derive the corresponding Hamiltonian. The numerical and analytical results are presented in Sec. \ref{sec:3}. In Sec \ref{sec:4} we discuss the effect of flux noise and in Sec. \ref{sec:5} we give our conclusion.

\section{Setup and Hamiltonian}
\label{sec:2}
In our setup four qubits are connected via a nonlinear coupler (see Fig. \ref{setup}). Here the qubits as well as the nonlinear coupler are realized by an inductive loop with inductivity $L_j$ interrupt by a Josephson junction with capacity $C_j$ and critical current $I_j^{(c)}$, namely a flux qubit. The junction represents the nonlinear ingredient of the system. Here $j \in \left\{1,2,3,4,c\right\}$. A crucial point is that the couplers plasma frequency has to be chosen higher than the qubit ones to avoid transitions between coupler energy levels. The corresponding quantum variables are the quantized fluxes of the five loops $\hat \Phi_j$. The Hamiltonian describing Fig.\ref{setup} can be obtained by standard circuit quantization. We split the Hamiltonian in three parts \cite{devoret1995quantum}, 
\begin{align}
\hat H = \sum_{j=1}^4\hat H_{j} + \hat H_c + \hat H_{\rm int},
\label{Hamilton}
\end{align}
the sum over the bare qubit Hamiltonians $\hat H_j$, the bare coupler Hamiltonian $\hat H_c$ and the interaction Hamiltonian $H_{\rm int}$. The qubit and the coupler Hamiltonian include a quadratic potential coming from the $LC$ part and a cosine contribution form the Josephson junction. The strength of the nonlinear term is determined by the ratio between the Josephson energy $E_J = \Phi_0 I_j^{(c)}/2\pi$ and the inductive energy $E_{L_j} = (\Phi_0/2\pi)^2/L_j$, where $\Phi_0$ denotes the flux quantum.
More interest should be paid into the interaction part, which arises from an induced external flux from qubit to the coupler and vice versa. Further we operate the coupler as well as the qubits at (or close to) the flux degeneracy point.
The different parts of \eqref{Hamilton} can be written in unitless parameters (for a detailed derivation see App. \ref{app1})
\begin{align}
\hat H_j &= E_{L_j} \left(4\xi_j^2 \frac{\hat q_j^2}{2} + (1+\alpha_j^2)\frac{\hat \varphi_j^2}{2} + \beta_j \cos(\hat \varphi_j)\right) \label{Hamilton1}\\
\hat H_c &= E_{\tilde L_c} \left(4\xi_c^2 \frac{\hat q_c^2}{2} + \frac{\hat \varphi_c^2}{2} + \beta_c \cos(\hat \varphi_c)\right) \label{Hamilton2}\\
\hat H_{\rm int} &= E_{\tilde L_c} \left(\sum_{i<j}^4 \alpha_i\alpha_j \hat\varphi_i\hat\varphi_j+\sum_{j=1}^4 \alpha_j \hat\varphi_c\hat\varphi_j \right).\label{Hamilton3}
\end{align}
Here we rescaled the coupler impedance $\tilde L_{c} = L_c - \sum_{\rm j=1}^4 \alpha_j M_j$ to decouple equations, where $M_j$ denotes the mutual inductances of the $j$-th qubit and $\alpha_j = M_j/L_j$ is the dimensionless mutual inductance. Additionally we defined the parameter $\xi_j = 4\pi Z_j/R_Q$ with characteristic impedance $Z_j=\sqrt{L_j/C_j}$ and the resistance quantum $R_Q = h/e^2$. The quantized phases are given by $\hat \varphi_j = (2\pi/\Phi_0)\hat\Phi_j+\pi$ and $\hat q_j$ is the conjugated quantum variable. Note that we shifted the appearing phases by $\pi$ leading to the positive sign in front of the cosine part. For coupler and qubits the screening parameter is given by $\beta_c = 2\pi\tilde L_c I_c^{(c)}/\Phi_0$ and $\beta_j = 2\pi  L_j I_j^{(c)}/\Phi_0$, respectively. Here $I_j^{(c)}$ denotes the critical current of the junctions.

To write down the Hamiltonian in a qubit representation we need to project it into the two level subspaces with respect to the qubits. A standard way of doing this is to approximate the two wells of the flux qubit potential as shifted harmonic oscillators \cite{irish2005dynamics} and interpret the two persistent current states of the qubit as the lowest eigenstates of these symmetrically shifted oscillators \cite{jin2012strong}. This leads to 
\begin{align}
\hat \varphi_j \approx s_j \hat Z_j,
\label{qubit_model}
\end{align}
where $\hat Z_j$ denotes the Pauli spin operator in the persistent current basis. The factor $s_j \propto (1-\braket{0_+|0_-}^2)^{-1/2}$ accounts for the fact that the two shifted ground states are not orthogonal, meaning that $s_j$ would be unity if the overlap of these states was zero (see Supplement for further details).
Using this notation, the interaction Hamiltonian can be written as
\begin{align}
\hat H_{\rm int} = E_{\tilde L_c}  \left((\alpha s)^2\sum_{i,j=1}^4 \hat Z_i\hat Z_j + \alpha s \sum_{i=1}^4  \hat Z_i \hat \varphi_c\right),
\label{Hamilton_int}
\end{align}
where we assume identical qubits ($\alpha_i s_i = \alpha s$ $\forall i$).
The first part induces two body local interactions between the qubits, which we call the direct coupling part and the second part describes the interaction between the qubits and the coupler modes, here referred to as indirect coupling. In commonly used coupler architectures one chooses parameters such that the direct coupling dominates and the two local interactions become strong. Here we want to use a different strategy, where we choose parameters such that the direct interaction part is rather small compared to the indirect coupling part, which gives rise to two but also higher local interactions. 
\begin{figure*}[t!]
\includegraphics[width=.99\textwidth]{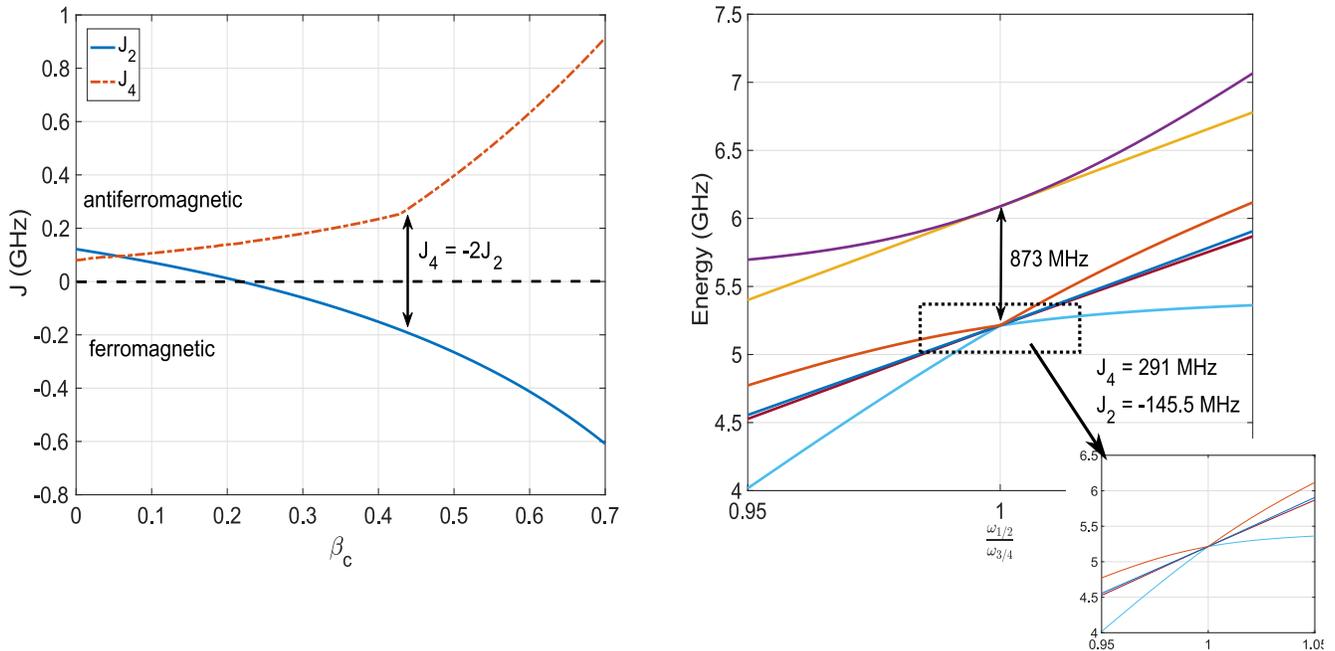}
\caption{\textit{Left: Coupling strength depending on the nonlinearity of the coupler for $E_{\tilde L_c} = 1$ THz, $\xi_c = 0.01$, $\xi_j = 0.05$, $\alpha_j = 0.05$ and $\beta_j = 1.1$. This corresponds to the physical qubit and coupler parameters $L_j = 817$ pH, $C_j=77$ fF and $L_c=170$ pH, $C_c=407$ fF, respectively with a mutual inductance of $M=40$ pH. Right: Numerically determined spectrum of the two excitation subspace for $E_j/E_{\rm \tilde L_c} = 0.2$. One sees exactly the spectrum theoretically expected at this specific point and the corresponding coupling strenghts are $J_4 = 291$ MHz, and $J_2 = -145.5$ GHz. The $x$-axis denotes the ratio between the frequency of qubit 1/2 and qubit 3/4, where we choose $\omega_1$ and $\omega_2$ constant and vary the frequency of the third and fourth qubit at the same time but equally.}}
\label{coupling}
\end{figure*}

To see how these higher order local interactions are indicated by the indirect interaction part we use a prominent tool from many body physics, the Schrieffer-Wolff transformation (SWT) \cite{schrieffer2002relation,bravyi2011schrieffer}. We choose the SWT since it produces physically transparent analytical expressions for the induced interactions arising from the indirect coupling part of \eqref{Hamilton_int}. Other than the simplest form of the Born-Oppenheimer approximation \cite{born1927zur}, it does not rely on the separation of classical frequencies and effective potentials. It is thus applicable even if transitions are vertical in the coordinate and if in the deep nonlinear regime classical frequencies are ill-defined (worse results for symmetric coupler potential in \cite{kafri2017tunable}). The SWT assumes that there are no transitions between different coupler levels, but includes corrections of the low energy subspace due to the existence of higher levels. Hence with the SWT it is possible to perturbatively write down an effective Hamiltonian in this low energy subspace
\begin{align}
\hat H_{\rm eff} = \hat P_0 \hat H_0 \hat P_0 + \epsilon \hat P_0 \hat V \hat P_0 + \sum_{n=2}^{\infty} \epsilon^n \hat H_{\rm eff,n}.
\label{SW_effective}
\end{align}  
where $\hat H_0 = \hat H_c + \sum_j \hat H_j$
is the unperturbed Hamiltonian, $\hat V = \hat H_{\rm int}$ is the perturbation, here the interaction, $\epsilon = \alpha s$ is a small parameter and $\hat P_0$ projects the Hamiltonian into the low energy subspace (coupler ground state).
Every order of the effective Hamiltonian leads to higher order local interactions between the qubits, i.e. in general the $k$-th perturbative term contains induced interactions up to $k$-th order. Truncation at fourth order therefore includes fourth order local interactions, such that the effective Hamiltonian has the form
\begin{align}
\begin{split}
\hat H \approx \hat P_0 \hat H_0 \hat P_0 +  J_2 \sum_{i<j}\hat Z_i \hat Z_j + J_4  \hat Z_1 \hat Z_2 \hat Z_3 \hat Z_4.
\label{4body_theory}
\end{split}
\end{align}
The coupling strengths $J_j$ are given by the prefactors generated by the SWT. Note that $J_2$ additionally includes direct interactions arising from the second term in \eqref{SW_effective}. In general the SW expansion also gives rise to single $\hat Z$-rotations and three body terms, but since we choose a symmetric coupler potential these terms are negligible, as we will argue in the following.

The physical principle behind these indirect interactions can be understood with the language of virtual excitations. E.g. the second order describes deexcitation of an excited qubit resulting in a virtual excitation of the coupler, which deexcites again and in turn excites the same or another qubit. Such processes leave the coupler in the ground state, but result in higher order qubit interactions. These virtual processes can be thought to occur only within the Heisenberg energy-time uncertainty. Fourth order processes in the same manner lead to four local interactions as illustrated in Fig. \ref{setup}. For the first and third order, there are no such processes, where the coupler ends up in the ground state, hence they can be neglected in \eqref{4body_theory} (2-3 orders of magnitude smaller as can be followed from the Hamiltonian and App. \ref{app3}).

\section{Results} 
\label{sec:3}

\subsection{Numerical results}
Since our qubit modeling using \eqref{qubit_model} is not very accurate for qubit nonlinearities only slightly larger than unity, we solve the system numerically and study the resulting spectrum. Here we evolve the bare qubit and coupler Hamiltonian in harmonic oscillator modes using about $50$ oscillator states, then project the interaction parts into the low energy subspace and determine the resulting spectrum numerically. The corresponding coupling strenghts can be extracted out of the spectrum by the distance of certain energy levels. In more detail we looked at the two-excitation subspace of the spectrum, which is also zoomed in on the right of Fig \ref{coupling}. Within this subspace, the distance between the different lines at the point where all frequencies are equal ($\frac{\omega_{1/2}}{\omega_{2/3}} = 1$) can be calculated analytically. These distances depend on $J_2$ and $J_4$, hence it is possible to translate the resulting spectrum into coupling strengths. 

The results for a device with realizable qubit and coupler parameters are shown in Fig. \ref{coupling}. All the contributions from the indirect coupling term increase with the coupler nonlinearity $\beta_c$. We see that for $\beta_c \approx 0$ the two local interactions are dominated by the antiferromagnetic direct coupling part and with increasing nonlinearity get more and more dominated by the ferromagnetic contribution from the indirect part. This results in a change of the nature of the interactions from antiferromagnetic to ferromagnetic at around $\beta_c = 0.2$. The four local interactions on the other hand are antiferromagnetic for all nonlinearities, since they only arise from the indirect coupling. Also we observe that $J_2$ and $J_4$ have a crossing point at around $\beta_c = 0.05$. For higher nonlinearites $|J_4|$ is larger than $|J_2|$. Both coupling strenghts increase with increasing $\beta_c$, but for the chosen parameters at $\beta_c$ around $0.7$ the energy levels of the coupler ground and coupler excited subspace start to mix, such that we can no longer use the setup to mimic the spectrum of the general Ising Hamiltonian including four local interactions. This is the reason why the results in Fig. \ref{coupling} are restricted to $\beta_c < 0.7$ (see App. \ref{app5}). 

A well distinguishable point in the spectrum is $J_4 = -2J_2$. In the two excitation subspace of the generalized four qubit Ising Hamiltonian including fourth order interactions, one observes three different energy levels, a non-degenerated, a twice degenerated and a three times degenerated one. At the specific point $J_4 = -2J_2$ this behavior changes and only two different energies are left over, a twice and a four times degenerated. Our results indicate that this point is at $\beta_c = 0.43$ and the numerically calculated spectrum for this specific nonlinearity is shown in Fig \ref{coupling} (right). We see exactly the theoretically expected behavior of the spectrum. For equal qubit frequencies the spectrum only shows two different energy levels, one twice and one four times degenerate. The distance between these two energy levels is $3J_4$. For the chosen parameters and a realistic $E_{\tilde L_c}$ of about $1$ THz, we observe a coupling strenght of $J_4 = 291$ MHz and $J_2 = -145.5$ MHz. By increasing $\beta_c$, the four body interactions can be tuned close to the GHz range. This is the largest predicted four local interaction strength in a superconducting qubit architecture without ancilla qubits, to the best of our knowledge. The coupling could also be made to be tunable by using a flux qubit architecture with tunable nonlinearity, i.e. a tunable rf-SQUID \cite{castellano2010deep} instead of the rf-SQUID coupler.
\begin{figure}
\includegraphics[width=.45\textwidth]{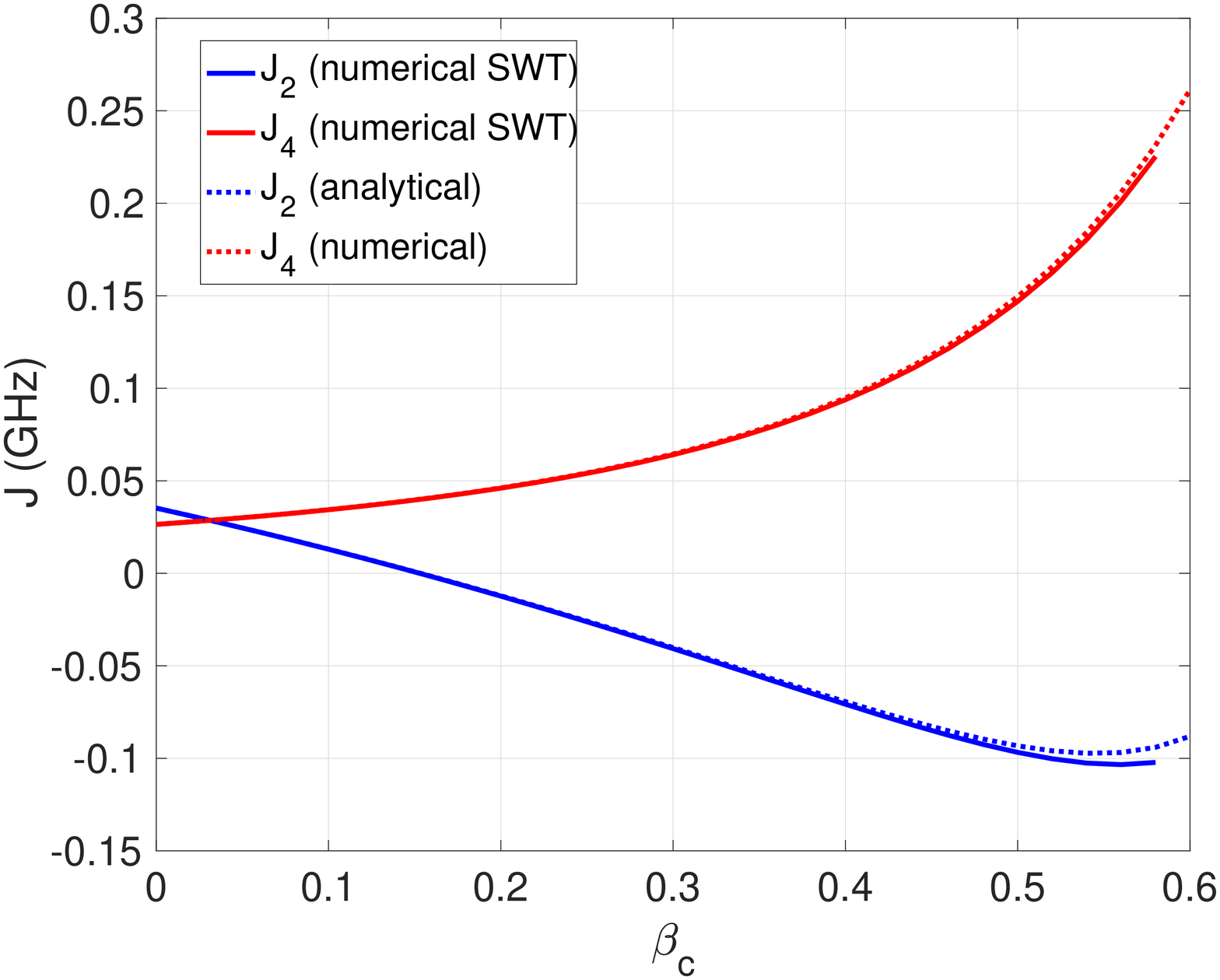}
\caption{\textit{Comparison between the coupling strengths for the fully analyical (dotted) and the numerical SWT for the same parameters as in Fig. \ref{coupling}}.}
\label{comparison}
\end{figure}
For the sense of completeness, we acknowledge that there is a four body flux qubit coupler proposal with comparable coupling strengths by Andrew J. Kerman \cite{kerman2018design}. In contrast to our setup they use a more complicated coupler setup, combining two devices, to realize the four local interactions, but therefore they are able to get rid of the two local interactions. Additionally there also exists recent a proposal by Melanson,  Martinez, Bedkihal and Lupascu to realize three local interactions using an additional twisted coupler \cite{melanson2019tunable}

\subsection{Analytical Results}
To get analytical results we calculate the SWT up to fourth order. This can be done in two different ways, the fully analytical way where we truncate the cosine of the coupler potential after fourth order and the numerical SWT where we keep the full cosine and need to calculate the respective coupling strengths resulting out of the SWT numerically. In both cases the qubit potentials are approximated via two shifted harmonic oscillators as explained in the last section and in more detail in App. \ref{app2}. Especially the fully analytical approach leads to the calculation of nasty commutators and long expressions for the coupling strengths. Therefore the full calculations can be found in App. \ref{app3} and we only talk about the results here. 

Although analytics qualitatively give the right behavior of the different coupling strengths, the actual values found by analytics differ from the numerical ones. As mentioned before, the reason is the rather small barrier of the qubit potential, making the validity of the shifted harmonic oscillator approximation we used to model the qubits not reliable. This leads to a wrong prefactor $s_j$ (see Eq. \eqref{qubit_model}) arising from the analytical qubit subspace projection. Also the larger $\beta_c$ gets, the more important higher orders of the SWT become. In Fig. \ref{comparison} we show the results for the fully analyitcal SWT and for the numrical SWT. We see that they show the right qualitative behavior and order of magnitude of the coupling strenghts, but the actual values differ from the ones found via numerics. E.g. for the specific point $\beta_c = 0.43$ chosen in Fig. \ref{coupling}, analytics predict $J_4 =\approx 112$ MHz and $J_2 \approx -80$ MHz. Another thing we see is that for $\beta_c \approx 0.55$ the two body coupling seems to have a minimum and then changes slope. This can be explained by the fact that larger $\beta_c$ leads to a smaller gap and at some point higher orders of the SWT are important. This means that we would not see this minimum but rather the behavior found in numerics when we start adding higher orders. Also higher orders of the couplers cosine potential part get more and more influence for higher $\beta_c$, explaining the slight differences between the two analyitcal approaches for large $\beta_c$.

\section{Flux noise effects}
\label{sec:4} 
 
 \begin{figure}
\includegraphics[width=.45\textwidth]{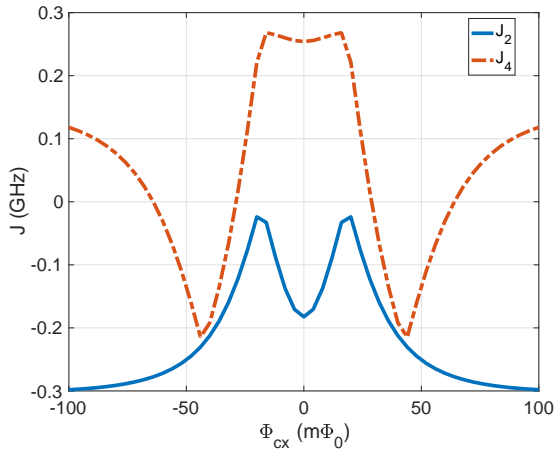}
\caption{\textit{Variation of the coupling strengths for a small external flux applied to the coupler loop. System parameters are the same as in Fig. \ref{coupling}}.}
\label{flux_noise}
\end{figure} 
So far we assumed a perfect coupler with four identical qubits and no external noise. However, in real flux qubit experiments a crucial effect is flux noise, which is present in all superconducting circuits. In the Hamiltonian \eqref{Hamilton1} and \eqref{Hamilton2} flux noise can be described as an additional external flux $\Phi_{jx}$ on the qubit and coupler loops, respectively. An external flux on the qubit loops induces a small tilt of the double well potential driving it slightly away from the flux degeneracy point. However, this effect just adds a small $\hat Z_j$ contribution on the qubit Hamiltonians. The most important influence of flux noise is the external noise applied to the coupler loop. This can significantly change the respective coupling strengths. In Fig. \ref{flux_noise} we show $J_2$ and $J_4$ under the influence of an external current on the coupler loop. We see that small flux variations do not change the four local interaction strength, indicating a magic point for flux noise at $\Phi_{cx}=\Phi_0/2$. Such a point arises when first order corrections of flux fluctuations vanish due to symmetry properties \cite{amin2005flux} when the first order dephasing rate induced by the environment (Fermis Golden rule contribution) vanishes. Only the two local interactions are affected. However, the two local interactions become smaller meaning we can apply an external flux to discriminate the two local interactions, leading to $J_4$ in the GHz range and $J_2$ about two orders of magnitude smaller. Driving the system slightly away from the degeneracy point also adds linear $\hat Z_j$ corrections to the qubits. However, for the chosen parameters the $\hat Z_j$ corrections can be estimated from the Hamiltonian to be approximately $10$ GHz per flux quantum, which results in $J_1 \approx 3$ MHz for the sweet spot shown in Fig. \ref{flux_noise}. Note that the flux offset range of Fig. \ref{flux_noise} is chosen orders of magnitude larger than typical flux offsets, to visualize the appearance of the plateau in the four local interactions.

\begin{figure}
\includegraphics[width=.45\textwidth]{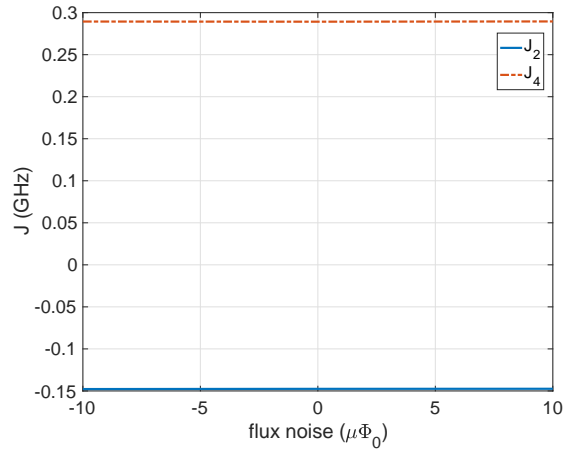}
\caption{\textit{Coupling strength of our setup depending on a flux offset induced in the coupler and all four qubits}.}
\label{flux_noise_all}
\end{figure}

Note that the range of flux offset shown in Fig. \ref{flux_noise} is way above flux offsets which are induced by typical noise sources, it was just chosen such that the appearance of a magic point is more visible. The most present noise in superconducting flux qubit architectures is 1/f flux noise, getting his name from the respective power spectral density being proportional to the reciprocal frequency \cite{RevModPhys.86.361}. However, typical flux noise amplitudes are around 1-10 $\mu \Phi_0$ \cite{PhysRevLett.99.187006}, meaning the induced flux offsets are also in this range. Since the whole circuit presented in the main text would be build on one chip, meaning it is surrounded by the same electrical environment, we assume a constant noise offset at both the coupler and all four qubits. Fig. \ref{flux_noise_all} shows the resulting coupling strength depending on the noise offset on coupler and all four qubits. We see that for typical noise amplitudes the change in coupling strength is not significant, meaning that the circuit is very robust against the usual flux noise appearing in superconducting architectures.

\begin{figure}
\begin{center}
\includegraphics[width=.45\textwidth]{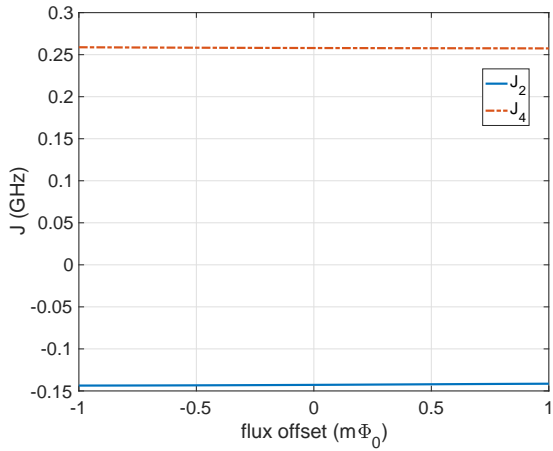}
\caption{\textit{Coupling strength of our setup depending on the flux offset in the coupler with the additional fixed flux offsets on the qubits $\Phi_{c,1}= 1$ m$\Phi_0$, $\Phi_{c,2}= 1.5$ m$\Phi_0$, $\Phi_{c,3}= -2.1$ m$\Phi_0$ and $\Phi_{c,4}= 3$ m$\Phi_0$.}}
\label{flux_offset_all}
\end{center}
\end{figure}

Another effect which is important to study is the effect of flux offsets in the controls, meaning different flux offsets acting on the four qubits. This leads to the four qubits being at four different bias points and symmetry effects such as the plateau seen in Fig. 3 of the main text could be no longer present. In real experiments there are two reasons for such flux offsets. On the one hand flux 1/f noise which leads to long term drifts because of non-vanishing expectation values. This value is obtained by the integrated power spectral density and standard values are in between $10$ and $100$ $\mu\Phi_0$. On the other hand such control offsets can be caused by resolution itself. With arbitrary waveform generators and a resolution of at least 10 bits with a rang split over a couple of flux quantum this leads to a rather high offset of about $1$ m$\Phi_0$. To prove that these effects do not affect our system dramatically, we choose four random flux offsets in the m$\Phi_0$ range and study the variation of the coupling strenghts depending on a variation of the coupler flux offset in the same range. The result is shown in Fig. \ref{flux_offset_all} and we again see that for typical values appearing in experiment, the effect on the coupling strength is not crucial. The overall value is slightly changed, but it stays constant over the whole range of coupler offsets, meaning it is not more susceptible to noise away from the symmetric bias case.

Another effect that appears in real experiments, especially when using artificial atoms as qubits are fabrication errors leading to uncertainties in the system parameters. This is studied in detail in App. \ref{app6} and we show that our setup is rather robust against these.

\section{Conclusion}
\label{sec:5}
In conclusion we have shown that the coupling architecture presented in Fig. \ref{setup} can exhibit large effective four body local interactions in the deep nonlinear regime. With suitable realistic  parameters they are even larger than the two body local interactions and are in the GHz range. To our knowledge these are (together with \cite{kerman2018design}) the strongest four body local interactions ever predicted in an architecture without additional ancilla qubits. Building such a device could yield a strong improvement of the applicability of quantum annealers. Also it would be possible to use a slightly different qubit arrangement or a twist in the coupler (comparable to \cite{melanson2019tunable}) to change the sign of the two and four local interactions to directly implement ferromagnetic four body interactions, which are often desired \cite{lechner2015quantum}. Such an additional coupler can also be used to completely get rid of the two local interaction as done in \cite{melanson2019tunable}, leading to a bare four body coupler.

The presented architecture could be used to build more efficient adiabatic quantum computers on the platform of superconducting flux qubits and exhibit the limits of currently used architectures (e.g. D-wave). Furthermore as discussed in the beginning, the realization of a four body coupler could be the next step towards the quantum technological implementation of a universal AQC.

\begin{acknowledgments}
We would like to thank Adrian Lupa\c{s}cu, Denis Melanson, Andrew J. Kerman and Simon J\"ager for fruitful discussions that certainly increased the quality of this work.

The research is based
upon work (partially) supported by the Office
of the Director of National Intelligence (ODNI),
Intelligence Advanced Research Projects Activity
(IARPA), via the U.S. Army Research Office
contract W911NF-17-C-0050. The views
and conclusions contained herein are those of
the authors and should not be interpreted as
necessarily representing the official policies or
endorsements, either expressed or implied, of
the ODNI, IARPA, or the U.S. Government.
The U.S. Government is authorized to reproduce
and distribute reprints for Governmental purposes
notwithstanding any copyright annotation
thereon.
\end{acknowledgments}

\begin{appendix}
\section{Circuit Hamiltonian}
\label{app1}

In this section we show how to get from Kirchhoffs laws to the Hamiltonian \eqref{app:Hamiltonian_circuit} of the circuit shown in Fig. \ref{setup} using circuit quantization. The section mostly recaps calculations that can be found in \cite{kafri2017tunable}, but for the sense of completeness we also show them here. 
In our setup we inductively couple four superconducting flux qubits using a coupling loop, realized by an additional flux qubit with higher plasma frequency as the four qubits. Kirchhoffs laws and Josephsons equations give the current equations of the system
\begin{align}
C \ddot \Phi_c +I_c^{(c)} \sin(2\pi \Phi_c/\Phi_0) - I_{L,c} &= 0 \label{app:Kirchhoff1} \\
I_j - I_j^* &=0 \hspace{1cm} (1 \leq j \leq k).\label{app:Kirchhoff2}
\end{align}
For the first equation, $\Phi_c$ denotes the flux across the coupler's Josephson junction (and capacitor), $I_{L,c}$ denotes the current through the couplers inductor, and $\Phi_0 = h/(2e)$ is the flux quantum. The second equation simply states that the current $I_j$ through $j$-th inductor is equal to the current $I_j^*$ flowing through the rest of the qubit circuit. We just leave the factors $I_j^*$ like this, since we will see that they do not give a contribution to the interaction part and later lead to the usual flux qubit Hamiltonian \cite{devoret1995quantum}. The inductive and flux quantization relationships can be combined into
\begin{align}
L_c I_{L,c} + \sum_{j=1}^k M_j I_j &= \Phi_{L,c}\label{app:quantize1} \\
L_j I_j + M_j I_{L,c} &= \Phi_j\label{app:quantize2} \\
\Phi_{L,c} = \Phi_{cx} - \Phi_c\label{app:quantize3},
\end{align}
where $\Phi_{cx}$ is the external flux applied to the coupler loop, $\Phi_j$ is the flux across the $j$-th junction, $L_j$ is the $j$-th qubit self inductance and $M_j$ is the mutual inductance between the $j$th qubit and the coupler. With equations \eqref{app:quantize1}-\eqref{app:quantize3} it is possible to rewrite equations \eqref{app:Kirchhoff1} and \eqref{app:Kirchhoff2} in flux variables
\begin{align}
C\ddot\Phi_c I_c^{(c)}\sin(2\pi \Phi_c/\phi_0) + \frac{\Phi_c-\Phi_{cx}+\sum_{i=j}^4 \alpha_i \Phi_j}{\tilde L_c} &=0 \label{app:Lagrange1}\\
\frac{\Phi_j}{L_j} + \alpha_j \left(\Phi_c - \Phi_{cx}+\sum_{k=1}^4 \alpha_k\Phi_k \right) - I_j^* = 0,\label{app:Lagrange2}
\end{align}
with dimensionless mutual inductance $\alpha_j = M_j/L_j$ and rescaled coupler impedance $\tilde L_c = L_c - \sum_{j=1}^4 \alpha_j M_j$. These equations of motion represent the Euler-Lagrange equation, resulting from the Lagrange function of the system. Now one can apply circuit quantization to find the corresponding Hamiltonian. From \eqref{app:Lagrange1} and \eqref{app:Lagrange2} we know the Lagrangian, which can be used to define the adjoint variable to the flux and write down a quantized version of the system Hamiltonian using the Legendre transformation. This leads to the Hamiltonian

\begin{align}
\begin{split}
\hat H &= \frac{\hat Q^2}{2C} - E_{J_c} \cos(2\pi \hat\Phi_c/\Phi_0) \\  &\hspace{1cm}+ \frac{\left(\hat \Phi_c - \Phi_{cx} + \sum_{j=1}^k \alpha_j \hat \Phi_j\right)^2}{2\tilde L_{c}} + \sum_{j=1}^k \hat H_j. 
\label{app:Hamiltonian_circuit}
\end{split}
\end{align}

Here $\hat H_j$ denotes the Hamiltonian for qubit $j$ in the absence of the coupler (i.e in the limit $\alpha_j \longrightarrow 0$). Here $\hat Q_c$ is the canonical conjugate to $\hat \Phi_c$ satisfying $\left[\hat \Phi_c,\hat Q_c\right] = i\hbar$, and the coupler's Josephson energy is $E_{J_c} = \Phi_0 I_c^{(c)}/2\pi$. 

The Hamiltonian can be rewritten in unitless parameters
\begin{align}
\hat H &= E_{\tilde L_c} \left(4\zeta_c^2 \frac{\hat q_v^2}{2} + \frac{(\hat \varphi_c-\varphi_x)^2}{2} + \beta_c \cos(\hat \varphi_c)\right) +\sum_{i=1}^4 \hat  H_j\\
\hat H_j &=  E_{\tilde L_j} \left(4\zeta_j^2 \frac{\hat q_j^2}{2} + \frac{(\hat \varphi_j-\varphi_x)^2}{2} + \beta_j \cos(\hat \varphi_j)\right),
\end{align}
with the following definitions:
\begin{align*}
E_{\tilde L_c} &= \frac{(\Phi_0/2\pi)^2}{\tilde L_c} 
\hspace{3cm}\xi_c = \frac{2\pi e}{\Phi_0}\sqrt{\frac{\tilde L_c}{C}} \\
\beta_c &= 2\pi \tilde L_c I_c^{(c)}/\Phi_0 = E_{J_c}/E_{\tilde L_c} 
\hspace{0.7cm}\hat q_c = \frac{\hat Q}{2e} \\
\hat \varphi_c &= \frac{2\pi}{\Phi_0} \hat \Phi_c + \pi 
\hspace{2.9cm}\varphi_{cx} = \frac{2\pi}{\Phi_0} \Phi_{cx} + \pi \\
\hat \varphi_j &= \frac{2\pi}{\Phi_0} \hat \Phi_j
\hspace{3.5cm}\hat \varphi_x = \varphi_{cx} - \sum_{j=1}^k \alpha_j \hat \varphi_j 
\end{align*}
Note that the phases $\hat \varphi$ is shifted by a factor $\pi$, such that the flux degeneracy point corresponds to $\varphi_{cx}= 0$.

\section{Projection into qubit subspace}
\label{app2}

Since we are interested in qubit interactions, we want to project the qubit part of the Hamiltonian into the subspace of the two lowest eigenstates of every included qubit (computational states). To do so we take a look at the qubit potential
\begin{align}
\hat U_j(\varphi_j) = \frac{1+\alpha_j^2}{2}\varphi_j^2 + \beta_j \cos(\varphi_j).
\end{align} 
In case of a flux qubit, the nonlinearity $\beta_j$ should be larger than one. This leads to a double well potential. The local maximum is located at $\varphi = 0$ and the two symmetric minima at $\varphi = \pm \varphi_p$. Now we approximate the two wells of the potential with two harmonic potentials, shifted by $\pm \varphi_p$ respectively . The equation that determines $\varphi_p$ reads
\begin{align}
0 = (1+\alpha_j) \varphi - \beta_c \sin(\varphi_p) \label{app:determine_minimum}, 
\end{align}
which can easily be solved numerically. To get a harmonic approximation, we evolve the respective potential well around $\pm \varphi_p$ up to second order
\begin{align}
U^+(\varphi) = c + U'(\varphi_p)(\varphi-\varphi_p) + U''(\varphi_p)(\varphi-\varphi_p)^2 \\
U^-(\varphi) = c + U'(-\varphi_p)(\varphi+\varphi_p) + U''(-\varphi_p)(\varphi+\varphi_p)^2.
\end{align}
The constant part can be ignored, and the first derivative vanishes, since $\varphi_p$ satisfies equation \eqref{app:determine_minimum}. Hence we get
\begin{align}
\begin{split}
U(\varphi) &\approx U^+(\varphi)+U^-(\varphi)\\
&=  \frac{1+\alpha_j - \beta_j\cos(\varphi_p)}{2}(\varphi-\varphi_p)^2 \\ &\hspace{0.2cm}+ \frac{1+\alpha_j-\beta_j\cos(\varphi_p)}{2}(\varphi+\varphi_p)^2.
\end{split}
\end{align}
To quantize the system we introduce the raising and lowering operator of the two shifted quadratic potentials
\begin{align}
\hat a_{\pm}^{\dag} \ket{N_{\pm}} &= \sqrt{N_{\pm} + 1}\ket{N_{\pm}+ 1}\\
\hat a_{\pm} \ket{N_{\pm}} &= \sqrt{N_{\pm}}\ket{N_{\pm}- 1},
\end{align}
where $\ket{N_{\pm}}$ are the Fock states of the respective shifted harmonic oscillator. For more details on the displaced harmonic oscillator basis we refer the reader to \cite{irish2005dynamics}. We want to restrict the basis to the two lowest energy levels (qubit basis). In the flux basis, which are the superpositions of the ground states $\ket{0_{\pm}}$ of the two wells 
\begin{align}
\ket{\tilde 0} &= \frac{1}{\sqrt{2}}\left(\ket{0_+}+\ket{0_-}\right) \\
\ket{\tilde 1} &= \frac{1}{\sqrt{2}}\left(\ket{0_+}-\ket{0_-}\right).
\end{align}
Here the two ground states $\ket{0_{\pm}}$ correspond to the persistent current states of the respective flux qubit.
The two states are othorgonal, but since $\braket{0_+|0_-}\neq 0$, we need to redefine an orthonormal qubit basis
\begin{align}
\ket{0} &= \frac{1}{\sqrt{2(1+\braket{0_+|0_-})}}\left(\ket{0_+}+\ket{0_-}\right) \\
\ket{1} &= \frac{1}{\sqrt{2(1-\braket{0_+|0_-})}}\left(\ket{0_+}-\ket{0_-}\right).
\end{align}
Using all the properties we wrote down in this section, we can translate the quantized phase into an operator only acting in the new defined qubit subspace
\begin{align}
\hat \varphi_j \longmapsto \frac{1}{\sqrt{2 m_j\omega_j(1-\braket{0_-|0_+}^2)}}\hat X_j,
\label{phi_quantized}
\end{align}
where the $m_j = 1/4\xi_j^2$ and $\omega_j =2\xi_j\sqrt{1+\alpha_j^2-\beta_c\cos\varphi_p}$ are the effective mass and frequency of the quadratic potential and $\hat X_j$ is the Pauli spin operator in the qubit basis. For simplification we addionally define the factor 
\begin{align*}
s_j = \frac{1}{\sqrt{2 m_j\omega_j(1-\braket{0_-|0_+}^2)}},
\end{align*}
which also appears in the main text. The overlap between the states in the displaced wells can be calculated by the formula \cite{irish2005dynamics}
\begin{align}
\braket{M_-|N_+} = \begin{cases}{\rm e}^{-\frac{\varphi_p^2}{2}}(-\varphi_p)^{M-N}\sqrt{N!/M!} L_N^{M-N}[\varphi_p^2] & M \geq N \\ {\rm e}^{-\frac{\varphi_p^2}{2}}(-\varphi_p)^{N-M}\sqrt{M!/N!} L_M^{N-M}[\varphi_p^2] & M < N \end{cases},
\end{align}
where $L_n^k$ are the generalized Laguerre polynomials. In the flux qubit literature it is more common to write down the Hamiltonian in the persistant current basis rather than the qubit basis, hence the flux is proportional to $Z_j$ instead of $X_j$ ($X_j$ $\mapsto$ $Z_j$) in the following and in the main text.

\begin{widetext} 
\section{Analytic approximation of the effective SW Hamiltonian}
\label{app3}

Here we will show how to get an analytical approximation for the SWT by expressing the harmonic part of the coupler using latter operators, and treating the cosine part as a perturbation. For this we truncate the coupler potential after $\mathcal{O}(\hat \varphi_c^4)$. The interaction part of the Hamiltonian then reads
\begin{align}
\hat H_{\rm int}/E_{\tilde L_c} = \frac{\beta_c}{24}\hat\varphi_c^4 + \alpha^2 \sum_{i<j}^4 \hat\varphi_i\hat\varphi_j - 2\varphi_{cx}\hat\varphi_c + \alpha \sum_{i=1}^4 \hat\varphi_c\hat\varphi_i + \varphi_{cx}\sum_{i=1}^4\hat\varphi_i.
\end{align}
For simplicity we chose identical qubits here. Now we assume a symmetric coupler potential $\varphi_{cx}=0$ and use \eqref{phi_quantized} to project the Hamiltonian into the qubit subspace
\begin{align}
\hat H_{\rm int}/E_{\tilde L_c} = \frac{\beta_c}{24}\hat\varphi_c^4 + \alpha^2 s^2 \sum_{i<j}^4 \hat Z_i \hat Z_j + \alpha s \sum_{i=1}^4 \hat\phi_c\hat Z_i.
\end{align}
The harmonic part of the coupler potential
\begin{align}
\hat H_c^{\rm harm} = 4E_{\tilde L_c}  \left(\xi_c^2 \frac{\hat q_c^2}{2} + \frac{1-\beta_c}{2}\hat\varphi_c^2 \right)
\end{align}
can be interpreted as a quantum harmonic oscillator with effective mass $m_c = 1/4E_{\tilde L_c}\xi_c^2$ and frequency $\omega_c = 2E_{\tilde L_c}\xi_c \sqrt{1-\beta_c}$. Note that the coupler nonlinearity $\beta_c$ is assumed to be smaller that one, since we want the coupler frequency to be higher than the qubit frequencies. With this we can define the position and momentum operator as
\begin{align}
\hat X_c &= \sqrt{m_c\omega_c } \hat\varphi_c \\
\hat P_c &= \frac{1}{\sqrt{m_c\omega_c}}\hat q_c
\end{align}
and rewrite the harmonic part as $\hat H_c^{\rm harm} = \frac{\omega_c}{2} (\hat X_c^2+\hat P_c^2)$. In the same manner we can define the annihilation and creation operator 
\begin{align}
\hat a_c &= \frac{1}{\sqrt{2}}\left(\hat X_c + i\hat P_c\right)\\
\hat a_c^{\dag} &= \frac{1}{\sqrt{2}}\left(\hat X_c - i\hat P_c\right)
\end{align}
and rewrite the quantized phase in terms of these operators
\begin{align}
\hat \varphi_c = \frac{1}{\sqrt{2m_c\omega_c}}\left(\hat a_c^{\dag} + \hat a_c \right).
\label{phi_c_ho}
\end{align}
Using equation \eqref{phi_c_ho} it is possible to rewrite the interaction Hamiltonian in terms of latter operators
\begin{align}
\hat H_{\rm int}/E_{\tilde L_c} = \frac{\beta_c}{24}\frac{1}{(2m_c\omega_c)^2}\left(\hat a_c^{\dag} + \hat a_c\right)^4 + \alpha^2 s^2 \sum_{i<j}^4 \hat Z_i \hat Z_j + \alpha s \sum_{i=1}^4 \frac{1}{\sqrt{2m_c\omega_c}}\left(\hat a^{\dag}_c + \hat a_c\right)\hat Z_i.
\label{int_ho}
\end{align} 
To get a better overview, we divide \eqref{int_ho} into three parts, the direct qubit-qubit coupling
\begin{align}
\hat H_{\rm int}^{QB,QB} = E_{\rm \tilde L_c} \alpha^2 s^2 \sum_{i<j} \hat Z_i \hat Z_j
\label{H_QBQB}
\end{align}
the indirect coupling part between qubits and coupler modes
\begin{align}
\hat H_{\rm int}^{QB,c} &= \frac{E_{\tilde L_c}}{\sqrt{2m_c\omega_c}} \alpha s\sum_{i=^1}^4 \left(\hat a_c^{\dag} + \hat a_c\right)\hat Z_i,
\label{H_QBc}
\end{align}
and the corrections arising from the fourth order cosine part
\begin{align}
\hat H_{\rm corr} = \frac{E_{\tilde L_c}\beta_c}{96m_c^2\omega_c^2} \left(\hat a_c^{\dag} + \hat a_c\right)^4.
\label{H_corr}
\end{align}
To simplify notation even more in the following, we define the appearing prefactors as follows:
\begin{align}
g^{\rm QB,c} &= \frac{E_{\tilde L_c}\alpha s}{\sqrt{m_c \omega_c}} \\
g^{\rm QB,QB} &= E_{\tilde L_c} \alpha^2 s^2 \\
K_{\rm corr} &= \frac{E_{\tilde L_c \beta_c}}{96m_c^2\omega_c^2}.
\end{align}
As mentioned in the main part of the paper, we want to perform the SWT under the assumption that the coupler frequency is higher than the respective qubit frequencies. Basically we have three different perturbative parts \eqref{H_QBQB}, \eqref{H_QBc} and \eqref{H_corr}, where $\hat H_{\rm int}^{QB,QB}$ just acts on the qubit subspace, hence simply gives a contribution in zeroth order. 
 
In a first step we have to calculate the even and odd contributions of the perturbative parts. Let's define $P_0 = \ket{0}\bra{0}$ and $Q_0 = 1-P_0 = \sum_{n=1}^{\infty} \ket{n}\bra{n}$ as the projection operator on the even and odd subspaces, respectively. Here $\ket{n}$ denotes the n-th Fock state of the harmonic coupler potential. The off-diagonal part of an operator $\hat X$ is then given by $\mathcal{O}(\hat X) = P_0 \hat X Q_0 + Q_0 \hat X P_0$ and the diagonal part by $\mathcal{D}(\hat X) = P_0\hat XP_0 + Q_0 \hat XQ_0$. Since $\hat H_{\rm int}^{\rm QB,QB}$ acts as identity on the coupler subspace it is completely diagonal. The other parts read
\begin{align}
\mathcal{O}(\hat a^{\dag} + \hat a) &= \eta_{01}^+ \\
\mathcal{O}\left((\hat a^{\dag}+\hat a)^4\right) &= \sqrt{4!}\eta_{04}^+ + 5\sqrt{2!}\eta_{02}^+ \\
\mathcal{D}(\hat a^{\dag} + \hat a) &= \sum_{n=1}^{\infty} \eta_{n,n+1}^+ \\
\mathcal{D}\left((\hat a^{\dag}+\hat a)^4\right) &= \sum_{n=1}^{\infty} \left(A_n^{(4)} \eta_{n,n+4}^+ + A_n^{(2)} \eta_{n,n+2}^+ + A_n^{(0)} \frac{\eta_{n,n}^+}{2}\right),
\end{align}
with $\eta_{k,l}^{\pm} = \ket{k}\bra{l} \pm \ket{l}\bra{k}$, $A_n^{(4)} = \sqrt{(n+4!/n!)}$ , $A_n^{(2)} = \sqrt{n^2(n+1)(n+2)} + \sqrt{(n+1)^3(n+2)} + \sqrt{(n+1)(n+2)^3} + \sqrt{(n+1)(n+2)(n+3)^2}$ and $A_n^{(0)} = 6(n^2+n)$. Additionally we calculate some useful commutators
\begin{align}
\left[\eta_{ij}^+,\eta_{kl}^+\right] &= \delta_{jk} \eta_{il}^{-} + \delta_{jl} \eta_{ik}^{-} + \delta_{ik} \eta_{jl}^{-} + \delta_{il}\eta_{jk}^{-} \label{comm1}\\
\left[\eta_{ij}^{-},\eta_{kl}^{-}\right] &= \delta_{jk} \eta_{il}^{-} - \delta_{jl} \eta_{ik}^{-} - \delta_{ik}\eta_{jl}^{-}+\delta_{il}\eta_{jk}^{-}\label{comm2} \\
\left[\eta_{ij}^{-},\eta_{kl}^{+}\right] &= \delta_{jk} \eta_{il}^{+}  + \delta_{jl}\eta_{ik}^{+} - \delta{ik}\eta_{jl}^{+} - \delta_{il}\eta_{jk}^{+}\label{comm3} \\
\left[\eta_{ij}^{+},\eta_{kl}^{-}\right] &= \delta_{jk} \eta_{il}^{+}  - \delta_{jl}\eta_{ik}^{+} + \delta{ik}\eta_{jl}^{+} - \delta_{il}\eta_{jk}^{+}\label{comm4}.
\end{align}

With this as a starting point we can calculate the different orders of the SW corrections to the effective Hamiltonian. The zeroth order of the effective Hamiltonian is just the unperturbed part projected into the coupler ground state subspace. The first order corrections are given by the diagonal projections of the perturbation, so in our case only the part $\hat H_{\rm int}^{\rm QB,QB}$ and the diagonal parts arising from $\hat H_{\rm corr}$, which are zero because $A_n^{0}(0)=0$. Hence a real calculation is only needed for the corrections of order larger than one. When calculated to a specific order the SWT finally gives an effective Hamiltonian acting only on the subspace of interest (here the coupler in ground state subspace), but including corrections coming from states not included in this subspace. The form of the effective Hamiltonian is given in Eq. (7) of the main text. In the following we calculate the different orders of the SWT, up to fourth order analytically.

\subsection{Second order effective Hamiltonian}
\label{app3_1}

First we calculate the first order of the generator $S$, that defines the SWT and is used to calculate the respective order of the effective Hamiltonian. The first order of $S$ is given by
\begin{align}
S_1 = \mathcal{L}(V_{\rm od}),
\end{align}
where we used the notation of Bravyi et al. \cite{bravyi2011schrieffer}, such that $V_{\rm od}$ just denotes all the off diagonal parts of the perturbation Hamiltonian and the linear map $\mathcal{L}$ is defined as
\begin{align}
\mathcal{L}(X) = \sum_{i,j} \frac{\braket{i|\mathcal{O}(X)|j}}{E_i-E_j}\ket{i}\bra{j},
\end{align}
where $\{\ket{i}\}$ is an orhtonormal eigenbasis of the unperturbed Hamiltonian. With this definition we can write down the expression for $S_1$
\begin{align}
S_1 = \sum_{i,j} \frac{\braket{i|V_{\rm od}^{\rm QB,c}|j}}{E_i-E_j}\ket{i}\bra{j} + \sum_{i,j} \frac{\braket{i|V_{\rm od}^{\rm corr}|j}}{E_i-E_j}\ket{i}\bra{j}. 
\end{align}
In our case the $\ket{i}$'s are the eigenstates of the bare coupler Hamiltonian (harmonic oscillator part). We need the following expressions to get $S_1$:
\begin{align}
\braket{i|\eta_{kl}^+|j} &= \delta_{ik}\delta_{jl} + \delta_{il}\delta_{kj} \\
\Rightarrow \braket{i|\eta_{10}^+|j} &= \frac{1}{E_1-E_0}\eta_{10}^- \\
\Rightarrow  \braket{i|\eta_{40}^+ + \eta_{20}^+|j} &= \frac{1}{E_4-E_0}\eta_{40}^- + \frac{1}{E_2-E_0}\eta_{20}^-,
\end{align}
such that we get
\begin{align}
S_1 &= \sum_{j=1}^4 \frac{g_j^{\rm QB,c} \hat Z_j}{E_1-E_0}\eta_{10}^- + K_{\rm corr} \left(\frac{\sqrt{4!}}{E_4-E_0} + \frac{5\sqrt{2!}}{E_2-E_0}\eta_{20}^-\right) \\
&= \sum_{j=1}^4 \gamma_j^{(1)} \hat Z_j \eta_{10}^- + \beta_1^{(1)} \eta_{40}^- + \beta_2^{(1)},
\end{align}
where $\gamma_j^{(1)} = g_j^{\rm QB,c}/(E_1-E_0)$, $\beta_1^{(1)} = \sqrt{4!}K_{\rm corr}/(E_4-E_0)$ and $\beta_2^{(2)} = 6\sqrt{2}K_{\rm corr}/(E_2-E_0)$.

The second order of the effective Hamiltonian is then given by
\begin{align}
H_{\rm eff,2} = b_1 P_0 \hat S_1(V_{\rm od}) P_0,
\end{align}
where we again adopt the notation of Barvyi et al. such that $\hat S_1(V_{\rm od}) = \left[S_1,V_{\rm od}\right]$. The prefactor $b_1$ is characterized by the equation
\begin{align}
b_{2n-1} = \frac{2(2^{2n}-1)B_{2n}}{(2n)!}
\end{align}
with Bernoulli numbers $B_n$. Using the commutation relations \eqref{comm1}-\eqref{comm4} and the fact that $P_0$ projects into the coupler ground state subspace - only terms proportional to $\eta_{00}$ give a contribution - we get
\begin{align}
\hat H_{\rm eff,2} = -\left[\sum_{i,j=1}^4 \alpha_i^{(1)}g_j^{\rm QB,c} \hat Z_i \hat Z_j + \beta_1^{(1)}\beta_1^{(0)} + \beta_2^{(1)}\beta_2^{(0)}\right],
\end{align}
with $\beta_1^{(0)} = \sqrt{4!}K_{\rm corr}$ and $\beta_2^{(0)} =  6\sqrt{2}K_{\rm corr}$.
\subsection{Third order effective Hamiltonian}
\label{app3_2}

The second order of the generator $S$ is given by
\begin{align}
S_2 = -\mathcal{L}\hat V_d(S_1),
\end{align}
where $V_d$ denotes the diagonal contributions of the perturbation Hamiltonian. In a first step we calculate $[V_d,S_1]$. Again with \eqref{comm1}-\eqref{comm4} we get 
\begin{align}
\begin{split}
[V_d,S_1] &= \sqrt{2}\sum_{i,j=1}^4 g_i^{\rm QB,c} \gamma_j^{(1)} \hat Z_i \hat Z_j\eta_{20}^+ + \sum_{j=1}^4 g_j^{\rm QB,c} \beta_1^{(1)}\left(\sqrt{5}\eta_{50}^+ + 2 \eta_{30}^+\right)\\&\hspace{0.5cm} + \sum_{j=1}^4 g_j^{\rm QB,c} \beta_2^{(1)} \hat Z_j \left(\sqrt{3}\eta_{30}^+ + \sqrt{2}\eta_{10}^+\right) +  K_{\rm corr} \left[\sum_{j=1}^4\gamma_j^{(1)}A_1^{(4)} \hat Z_j \eta_{50}^+ + \beta_1^{(1)} A_2^{(4)}\eta_{60}^+ \right.\\ &\hspace{0.5cm}\left.+ \beta_2^{(1)}A_4^{(4)}\eta_{80}^++ \sum_{j=1}^4\gamma_j^{(1)}A_1^{(2)}\eta_{30}^+ \beta_1^{(1)}A_2^{(2)}\eta_{40}^+ + \beta_2^{(1)}A_4^{(2)} \eta_{60}^+ \beta_2^{(1)}A_2^{(2)}\eta_{20}^+ + \sum_ {j=1}^4  \gamma_j^{(1)}A_1^{(0)}\hat Z_j \eta_{10}^+\right. \\ &\hspace{0.5cm}+\left. \beta_1^{(1)} A_4^{(0)} \eta_{40}^+\beta_2^{(1)}A_2^{(0)}\eta_{20}^+ \right]
\end{split}
\end{align}
The next order of the effective Hamiltonian is given by $H_{\rm eff,3} = b_1P_0\hat S_2(V_{\rm od})P_0$, so it is again sandwiched by projection operators onto the coupler ground state. $S_2$ is given by $-\mathcal{L}(V_d(S_1))$. $\mathcal{L}$ maps $\eta_{ij}^+$ to $\eta_{ij}^-$ and adds the respective energy prefactor $1/(E_i-E_j)$. Only terms proportional to $\eta_{00}$ will not be projected to zero by $P_0$. Looking at \eqref{comm1}-\eqref{comm4} we see that only commutators of $\eta$s with identical indices will give a contribution. In $V_{\rm od}$ the only appearing operators of this sort are $\eta_{10}$, $\eta_{20}$ and $\eta_{40}$. This means that we can ignore all other $\eta$ operators in the commutator $\hat V_d(S_1)$, since they don't give a contribution to $H_{\rm eff,3}$. Using this simplification, we get the following expression for the third order effective Hamiltonian
\begin{align}
\hat H_{\rm eff,3} &= \left[\sum_{i,j=1}^4 \left(\frac{\sqrt{2}\beta_2^{(0)}}{(E_2-E_0)}\gamma_i^{(1)}g_j^{\rm QB,c} + \frac{\sqrt{2}\beta_2^{(1)}}{E_1-E_0} g_{i}^{\rm QB,c}g_i^{\rm QB,c} + \frac{K_{\rm corr} A_1^{(0)}}{E_1-E_0}\alpha_i^{(1)}g_j^{\rm QB,c}\right)\hat Z_i \hat Z_j \right.\\
&\left.\hspace{0.5cm}+ K_{\rm corr} \hat H_{\rm shift}^{(3)}\right]
\end{align}
where $H_{\rm shift}^{(3)}$ adds an overall energy shift to the coupler ground state energy given by
\begin{align}
H_{\rm shift}^{(3)} = \sum_{j=1}^4\left(\frac{\beta_1^{(1)}\beta_1^{(0)}A_2^{(2)}}{E_4-E_0} + \frac{\beta_2^{(1)}\beta_2^{(0)}A_2^{(2)}}{E_2-E_0} + \frac{\beta_1^{(1)}\beta_1^{(0)}A_4^{(0)}}{E_4-E_0} + \frac{\beta_2^{(1)}\beta_2^{(0)}A_2^{(0)}}{E_2-E_0}\right).
\end{align}
Hence the third order effective Hamiltonian has two effects on the qubits. It leads to an overall energy shift given by $H_{\rm shift}^{(3)}$ and like the second order effective Hamiltonian induces two body local interactions. Therefore we have to calculate the next higher order and see if local interactions $k>2$ appear.

\subsection{Fourth order effective Hamiltonian}
\label{app3_3}

The third part of the generator is given by 
\begin{align}
S_3 = -\mathcal{L}\hat V_d(S_2) + a_2\mathcal{L}\hat S_1^2(V_{\rm od}).
\end{align}
with parameters
\begin{align}
a_n = \frac{2^n B_n}{n!}
\end{align}
We start with calculating  $\hat V_d S_2$. In the expression for $H_{\rm eff,4}$ the commutator of $S_3$ with $V_{\rm od}$ appears. This expression is again sandwiched by $P_0$ operators. In the same manner as in the last section we therefore only have to include terms of $S_3$ proportional to $\eta_{10}$, $\eta_{40}$ or $\eta_{20}$. This leads to twelve different terms. The effective Hamiltonian is given by
\begin{align}
H_{\rm eff,4} = b_1 P_0\hat S_3(V_{\rm od}) P_0 + b_3 P_0\hat S_1^3(V_{\rm od}).
\end{align}
We split this Hamiltonian into three different parts
\begin{align}
H_{\rm eff,4} &= H_{\rm eff,4}^{(1)} + H_{\rm eff,4}^{(2)} + H_{\rm eff,4}^{(3)} \\
&= -b_1P_0\left[\mathcal{L}\hat V_d(S_2),V_{\rm od}\right]P_0 + b_1a_2P_0\left[\mathcal{L}\hat S_1^2(V{\rm od}),V_{\rm od}\right] + b_3P_0\hat S_1^3(V_{\rm od})P_0
\end{align}
For the first part we get:
\begin{align}
\begin{split}
H_{\rm eff,4}^{(1)}
&=-2b_1 K_{\rm corr}\sum_{i,j=1}^4\left(\frac{5\beta_1^{(1)}g_i^{\rm QB,c}g_j^{\rm QB,c}}{(E_5-E_0)(E_4-E_0)}+\frac{\sqrt{5}K_{\rm corr}A_1^{(4)}\gamma_i^{(1)}g_i^{\rm QB,c}}{(E_5-E_0)(E_4-E_0)}+\left(\frac{\sqrt{4}}{E_4-E_0}+\frac{\sqrt{3}}{E_3-E_0}\right)\frac{\sqrt{3}\beta_2^{(1)}g_i^{\rm Qb,c}g_j^{\rm QB,c}}{E_3-E_0}\right.\\ &\left.\hspace{0.5cm}+\frac{K_{\rm corr}A_1^{(2)}\gamma_i^{(1)}g_j^{\rm QB,c}}{E_3-E_0}\left(\frac{\sqrt{4}}{E_4-E_0}+\frac{\sqrt{3}}{(E_3-E_0)}\right) + \frac{2\beta_1^{(1)}g_i^{\rm QB,c}g_j^{\rm QB,c}}{E_3-E_0} \left(\frac{\sqrt{4}}{E_4-E_0}+\frac{\sqrt{3}}{E_2-E_0}\right) \right.\\ &\left.\hspace{0.5cm}+ \frac{2\beta_2^{(1)}\gamma_i^{1}g_j^{\rm QB,c}}{(E_2-E_0)^2} + \frac{\sqrt{2}K_{\rm corr}A_1^{(0)}\gamma_i^{(1)}\gamma_j^{(1)}}{E_2-E_0}+\frac{\sqrt{2}K_{\rm corr}A_2^{(2)}\gamma_i^{(1)}g_j^{\rm QB,c}}{(E_2-E_0)(E_4-E_0)}+\frac{\sqrt{2}K_{\rm corr} A_2^{(0)}\gamma_i^{(1)}g_j^{\rm QB,c}}{(E_2-E_0)^2}\right) \hat Z_i \hat Z_j \\
&\hspace{0.5cm}+ \sum_{j=1}^4\left(\frac{\sqrt{2}K_{\rm corr}\beta_2^{(1)}A_2^{(2)}g_j^{\rm QB,c}}{(E_2-E_0)(E_1-E_0)}+\frac{\sqrt{2}K_{\rm corr}\beta_2^{(1)}A_2^{(0)}g_j^{\rm QB,c}}{(E_2-E_0)(E_1-E_0)}+\frac{\sqrt{5}K_{\rm corr}\beta_1^{(1)}A_1^{(4)}g_j^{\rm QB,c}}{(E_5-E_0)(E_1-E_0)}+\frac{K_{\rm corr}A_1^{4}A_1^{(4)}\gamma_j^{(1)}}{(E_5-E_0)(E_1-E_0)} \right.\\&\left.\hspace{0.5cm}+ \frac{2K_{\rm corr}\beta_1^{(1)}A_1^{(2)}g_j^{\rm QB,c}}{(E_3-E_0)(E_2-E_0)}+\frac{\sqrt{3}K_{\rm corr}\beta_2^{(1)}A_1^{(2)}g_j^{\rm QB,c}}{(E_3-E_0)(E_1-E_0)}+\frac{K_{\rm corr}A_1^{(2)}A_1^{(2)}\gamma_j^{(1)}}{(E_3-E_0)(E_1-E_0)}+\frac{\sqrt{2}A_1^{(0)}\beta_2^{(1)}g_j^{\rm QB,c}}{(E_1-E_0)^2}\right.\\&\left.\hspace{0.5cm}+ \frac{K_{\rm corr}A_1^{(0)}A_1^{(0)}\gamma_j^{(1)}}{(E_1-E_0)^2}\right) \hat Z_j \\
&\hspace{0.5cm}+ \frac{K_{\rm corr}^2A_2^{(4)}\beta_2^{(1)}A_2^{(4)}}{(E_6-E_0)(E_2-E_0)} + \frac{K_{\rm corr}^2 \beta_2^{(1)}A_4^{(2)}A_2^{(2)}}{(E_2-E_0)(E_4-E_0)} + \frac{K_{\rm corr}^2\beta_2^{(1)} A_2^{(0)}A_2^{(2)}}{(E_2-E_0)(E_4-E_0)} + \frac{K_{\rm corr}^2 \beta_1^{(1)}A_2^{(2)}A_2^{(2)}}{(E_4-E_0)(E_2-E_0)} \\ &\hspace{0.5cm}+ \frac{K_{\rm corr}^2 \beta_1^{(1)}A_4^{(0)}A_2^{(2)}}{(E_4-E_0)(E_2-E_0)} + \frac{K_{\rm corr}^2  \beta_1^{(1)}A_2^{(4)}A_4^{(2)}}{(E_6-E_0)(E_4-E_0)} + \frac{K_{\rm corr}^2\beta_2^{(1)}A_4^{(2)}A_4^{(2)}}{(E_6-E_0)(E_4-E_0)} + \frac{K_{\rm corr}^2A_2{(0)}\beta_2^{(1)}A_2^{(2)}}{(E_2-E_0)^2} \\ &\hspace{0.5cm}+ \frac{K_{\rm corr} \beta_2^{(1)}A_2^{(0)}A_2^{(0)}}{(E_2-E_0)^2} + \frac{K_{\rm corr} \beta_1^{(1)}A_4^{(0)}A_2^{(2)}}{(E_4-E_0)^2} + \frac{K_{\rm corr} \beta_1^{(1)}A_4^{(0)}A_4^{(0)}}{(E_4-E_0)^2} + \frac{K_{\rm corr}^2\beta_1^{(1)}A_2^{(4)}A_2^{(4)}}{(E_6-E_0)(E_2-E_0)} + \frac{K_{\rm corr}^2 \beta_2^{(1)}A_4^{(4)}A_4^{(4)}}{(E_8-E_0)(E_4-E_0)} \\
&\hspace{0.5cm} + \sum_{i,j,k=1}^4 \frac{2 K_{\rm corr}\gamma_i^{(1)}g_j^{\rm QB,c}g_k^{\rm QB,c}}{(E_2-E_0)(E_1-E_0)} \hat Z_i \hat Z_j \hat Z_k
\end{split}
\end{align}
We see that a lot of two local coupling terms arise. Additionally we have single $\hat Z$ corrections, an overall energy shift and most important the last term leads to three local qubi-qubit interactions. Let's first calculate the other contributions to the effective Hamiltonian. The second part is given by
\begin{align}
\begin{split}
H_{\rm eff,4}^{(2)} &= 2b_1a_1\left[4\sum_{i,j,k,l=1}^3 \frac{g_i^{\rm QB,c}g_j^{\rm QB,c}g_k^{\rm QB,c}g_l^{\rm QB,c}}{(E_1-E_0)^4}\hat Z_i \hat Z_j \hat Z_k\hat Z_l + \sum_{i,j=1}^4\left(\frac{\gamma_i^{(1)}\gamma_j^{(1)}\beta_1^{(0)}\beta_1^{(0)}}{E_4-E_0} + \frac{\gamma_i^{(1)}\gamma_j^{(1)}\beta_2^{(0)}\beta_2^{(0)}}{E_2-E_0} \right.\right.\\&\left.\left.\hspace{0.5cm}	+ \frac{\beta_1^{(0)}\beta_1^{(1)}\gamma_i^{(1)}g_j^{\rm QB,c}}{E_4-E_0} + 2\beta_1^{(1)}\beta_1^{(0)}\gamma_i^{(1)}\gamma_j^{(1)} + \frac{\beta_2^{(0)}\beta_2^{(0)}\gamma_i^{(1)}g_j^{\rm QB,c}}{E_2-E_0} + 2\beta_2^{(1)}\beta_2^{(0)}\gamma_i^{(1)}\gamma_j^{(1)} + \frac{2\beta_2^{(1)}\beta_2^{(0)}\gamma_i^{(1)}g_j^{\rm QB,c}}{E_2-E_0} \right.\right.\\&\left.\left.\hspace{0.5cm}+ \beta_2^{(1)}\beta_2^{(0)}\beta_2^{(0)}\gamma_i^{(1)}\gamma_j^{(1)} +\beta_2^{(1)}\beta_2^{(1)}\gamma_i^{(1)}g_j^{\rm QB,c} + \frac{2\beta_1^{(1)}\beta_1^{0}\gamma_i^{(1)}g_j^{\rm QB,c}}{E_4-E_0} + \beta_1^{(1)}\beta_1^{(1)}\gamma_i^{(1)}g_j^{\rm QB,c} + \beta_1^{(1)}\beta_1^{(0)} \gamma_i^{(1)}\gamma_j^{(1)}\right) \hat Z_i\hat Z_j \right. \\&\left.\hspace{0.5cm}+\frac{\beta_2^{(1)}\beta_1^{(1)}\beta_2^{(0)}\beta_1^{(0)}}{E_4-E_0} + \frac{2\beta_1^{(1)}\beta_1^{(0)}\beta_2^{(1)}\beta_2^{(0)}}{E_2-E_0} + \frac{4\beta_1^{(1)}\beta_1^{(1)}\beta_1^{(0)}\beta_1^{(0)}}{E_4-E_0} + \frac{\beta_2^{(1)}\beta_2^{(1)}\beta_1^{(0)}\beta_1^{(0)}}{E_4-E_0} + \frac{\beta_1^{(1)}\beta_1^{(1)}\beta_2^{(0)}\beta_2^{(0)}}{E_2-E_0} \right. \\ &\left.\hspace{0.5cm}+ \frac{\beta_2^{(1)}\beta_1^{(0)}\beta_1^{1()}\beta_2^{(0)}}{E_2-E_0} + \frac{2\beta_1^{(1)}\beta_2^{(1)}\beta_2^{(0)}\beta_1^{(0)}}{E_4-E_0}\right] 
\end{split}
\end{align}

and the last part reads

\begin{align}
\begin{split}
H_{\rm eff,4}^{(3)} &= 2b_3\left[4\sum_{i,j,k,l=1}^4 \gamma_i^{(1)}\gamma_j^{(1)}\gamma_k^{(1)} g_l^{\rm QB,c} \hat Z_i\hat Z_j\hat Z_k\hat Z_l + \sum_{i,j=}^4 \left(\beta_1^{(0)}\beta_1^{(1)}\gamma_i^{(1)}\gamma_j^{(1)} + \beta_2^{(0)}\beta_2^{(1)}\gamma_i^{(1)}\gamma_j^{(1)}\right.\right. \\&\left.\left.\hspace{0.5cm}+\beta_1^{(1)}\beta_1^{(1)}\gamma_i^{(1)} g_j^{\rm QB,c} + 2\beta_1^{(0)}\beta_1^{(0)}\gamma_i^{(1)}\gamma_j^{(1)} + \beta_2^{(1)}\beta_2^{(1)}\gamma_i^{(1)}g_j^{\rm QB,c} + 2 \beta_2^{(0)} \beta_2^{(1)}\gamma_i^{(1)}\gamma_j^{(1)} + 2 \beta_2^{(1)}\beta_2^{(1)}\gamma_i^{(1)}g_j^{\rm QB,c}\right.\right. \\ &\left.\left.\hspace{0.5cm}+ \beta_2^{(0)}\beta_2^{(1)}\gamma_i^{(1)}\gamma_j^{(1)} + \beta_2^{(1)}\beta_2^{(1)}\gamma_i^{(1)}g_j^{\rm Qb,c}\right)\hat Z_i\hat Z_j + 2\beta_1^{(1)}\beta_1^{(0)}\beta_2^{(1)}\beta_2^{(1)} + \beta_2^{(1)}\beta_1^{(1)}\beta_2^{(0)}\beta_1^{(1)} 
+ \beta_2^{(1)}\beta_2^{(1)}\beta_1^{(0)}\beta_1^{(1)} \right.\\&\left.\hspace{0.5cm} 4\beta_2^{(1)}\beta_2^{(1)}\beta_2^{(0)}\beta_2^{(1)} + 4\beta_1^{(1)}\beta_1^{(1)}\beta_1^{(0)}\beta_1^{(1)} + \beta_1^{(1)}\beta_1^{(1)}\beta_2^{(0)}\beta_2^{(1)} + \beta_1^{(1)}\beta_2^{(1)}\beta_1^{(0)}\beta_2^{(1)} + 2\beta_1^{(1)}\beta_2^{(0)}\beta_2^{(0)}\beta_1^{(0)}\right].
\end{split}
\end{align}

Finally with all these results, the fourth order effective Hamiltonian acting only on the coupler ground state subspace can be written as
\begin{align}
H_{\rm eff} = \hat P_0 \hat H_0 \hat P_0 + \hat P_0 \hat V \hat P_0 + \sum_{n=2}^4 \hat H_{{\rm eff},n}.
\end{align}

\subsection{Coupling Strengths}
\label{app3_4}

All in all $H_{\rm eff,4}$ leads to $3$ and $4$ local qubit-qubit interactions. Anyways, there still are 2 local qubit interactions present and we want the higher ones, to give the leading effect. Therefore it is necessary to go to a regime where the 2 local interactions vanish or at least are smaller than the higher ones. Note that the sum $\sum_{i,j,k,l}$ also gives rise to 2 local interactions (e.g. if $i=j$ and $k=l$), so we also have to take them into account. 

To summarize the results, we want to give expressions for the different couplings. For simplification we assume that the qubit parameters are the same for all qubits. We define the different coupling strengths such that we can write the effective interaction Hamiltonian as:
\begin{align*}
H_{\rm int, eff} =  J_4  \hat \hat Z_1 \hat Z_2 \hat Z_3 \hat Z_4 +  J_3 \sum_{i< j< k}  \hat Z_i\hat Z_j\hat Z_k +  J_2 \sum_{i < k} \hat Z_i \hat Z_j + J_1 \sum_{i=1}^4 \hat Z_i,
\end{align*}
where the whole interaction Hamiltonian acts only on the $\ket{0}$ subspace of the coupler. The restriction of the sums comes from the fact that e.g $\hat Z_i \hat Z_j = \hat Z_j \hat Z_i$, hence we get an additional prefactor into the different coupling terms
\begin{align}
 &\sum_{i\neq j\neq k \neq l} \hat  \hat Z_i \hat Z_j \hat Z_k \hat Z_l + \sum_{i\neq j\neq k}  \hat Z_i \hat Z_j \hat Z_k +  \sum_{i \neq k} \hat Z_i \hat Z_j +\sum_{i=1}^4 \hat Z_i \\
 &= 4! \hat Z_i \hat Z_j \hat Z_k \hat Z_l + 3 ! \sum_{i< j< k}  \hat Z_i \hat Z_j \hat Z_k +  2! \sum_{i < k} \hat Z_i \hat Z_j +\sum_{i=1}^4 \hat Z_i.
 \label{app:prefactors}
\end{align}
The four body coupling strength is given by
\begin{align}
J_4 = 24\frac{g_{\rm QB,c}^4}{\Delta_{10}^3}, 
\end{align}
with $\Delta_{ij} = E_i-E_j$. The three body coupling strength is given by
\begin{align}
J_3 = -6 \frac{2 K_{\rm corr}  g_{\rm QB,c}^3}{\Delta_{20}\Delta_{10}^2}.
\end{align}
The expression for the two body interaction is a little more complicating
\begin{align}
\begin{split}
J_2/2 &= g_{\rm QB,QB}/2 - \frac{g_{\rm QB,c}^2}{\Delta_{10}} + \frac{12 K_{\rm corr} g_{\rm QB,c}^2}{\Delta_{20}\Delta_{10}} + \frac{12 K_{\rm corr} g_{\rm QB,c}^2}{\Delta_{10}\Delta_{20}} + \frac{12 K_{\rm corr} g_{\rm QB,c}^2}{\Delta_{10}^2} - \frac{10\sqrt{6}K_{\rm corr}^2 g_{\rm QB,c}^2}{\Delta_{50}\Delta_{40}^2} \\ &\hspace{0.5cm}- \frac{10\sqrt{6}K_{\rm corr}^2g_{\rm QB,c}^2}{\Delta_{50}\Delta_{40}\Delta_{10}} - \frac{12\sqrt{6}K_{\rm corr}^2g_{\rm QB,c}^2}{\Delta_{40}\Delta_{30}\Delta_{20}} - \frac{18\sqrt{2}K_{\rm corr}^2g_{\rm QB,c}^2}{\Delta_{30}^2\Delta_{20}} - \frac{20\sqrt{6}K_{\rm corr}^2 g_{\rm QB,c}^2}{\Delta_{40}\Delta_{30}\Delta_{10}} - \frac{30\sqrt{2}K_{\rm corr}^2g_{\rm QB,c}^2}{\Delta_{30}^2\Delta_ {10}} \\ &\hspace{0.5cm}  - \frac{8\sqrt{6}K_{\rm corr}^2g_{\rm QB,c}^2}{\Delta_{40}\Delta_{20}\Delta_{10}} - \frac{12\sqrt{2}K_{\rm corr}^2g_{\rm QB,c}^2}{\Delta_{20}^3} - \frac{12\sqrt{2}K_{\rm corr}^2 g_{\rm QB,c}}{\Delta_{20}\Delta_{10}^2} - \frac{28\sqrt{8}K_{\rm corr}^2g_{\rm QB,c} }{\Delta_{40}\Delta_{20}\Delta_{10}} - \frac{36\sqrt{2}K_{\rm corr}^2g_{\rm QB,c}^2}{\Delta_{20}^2\Delta_{10}} \\ &\hspace{0.5cm}+ \frac{8K_{\rm corr}^2g_{\rm QB,c}^2}{\Delta_{40}\Delta_{10}^2} + \frac{24 K_{\rm corr}^2g_{\rm QB,c}^2}{\Delta_{20}^3} + \frac{8 K_{\rm corr}^2 g_{\rm QB,c}^2}{\Delta_{40}^2\Delta_{10}} + \frac{16 K_{\rm corr}^2 g_{\rm QB,c}^2}{\Delta_{10}^3} + \frac{48K_{\rm corr}^2 g_{\rm QB,c}}{\Delta_{20}^3} + \frac{48K_{\rm corr}^2 g_{\rm QB,c}^2}{\Delta_{20}\Delta_{10}^2} \\ &\hspace{0.5cm}+ \frac{48K_{\rm corr}^2 g_{\rm QB,c}^2}{\Delta_{20}^2\Delta_{10}} + \frac{144\sqrt{2} K_{\rm corr}^2 g_{\rm QB,c}}{\Delta_{10}^2\Delta_{20}} + \frac{24 K_{\rm corr}^2 g_{\rm QB,c}^2}{\Delta_{10}\Delta_{20}^2} + \frac{48 K_{\rm corr}^2 g_{\rm QB,c}^2}{\Delta_{40}\Delta_{10}\Delta_{20}} + \frac{24 K_{\rm corr}^2 g_{\rm QB,c}}{\Delta_{10}\Delta_{20}^2}  + \frac{24 K_{\rm corr}^2g_{\rm QB,c}^2}{\Delta_{10}^2\Delta_{20}} \\ &\hspace{0.5cm}-\frac{2K_{\rm corr}^2 g_{\rm QB,c}^2}{\Delta_{40}\Delta_{10}^2} - \frac{6 K_{\rm corr}^2 g_{\rm QB,c}^2}{\Delta_{10}^2\Delta_{20}}
-\frac{2 K_{\rm corr}^2 g_{\rm QB,c}^2}{\Delta_{10}\Delta_{40}^2} - \frac{4K_{\rm corr}^2 g_{\rm QB,c}^2}{\Delta_{10}^2\Delta_{40}} - \frac{6K_{\rm corr}^2g_{\rm QB,c}^2}{\Delta_{10}\Delta_{20}^2} - \frac{12 K_{\rm corr}^2 g_{\rm QB,c}^2}{\Delta_{10}^2\Delta_{20}}\\ &\hspace{0.5cm}-\frac{12 K_{\rm corr}^2 g_{\rm QB,c}^2}{\Delta_{10}^2\Delta_{20}} - \frac{6K_{\rm corr}^2 g_{\rm QB,c}^2}{\Delta_{10}^2\Delta_{20}} - \frac{6K_{\rm corr}^2 g_{\rm QB,c}^2}{\Delta_{10}\Delta_{20}^2} - \frac{2K_{\rm corr}^2g_{\rm QB,c}^2}{\Delta_{10}\Delta_{20}^2} + \frac{20g_{\rm QB,c}^4}{\Delta_{10}^3} \end{split}
\end{align}
as well as the $\hat Z$ corrections
\begin{align}
\begin{split}
J_1 &=   -\frac{72\sqrt{3}K_{\rm corr}^3 g_{\rm QB,c}}{\Delta_{20}^2\Delta_{10}} - \frac{432 K_{\rm corr}^3 g_{\rm QB,c}}{\Delta_{20}^2\Delta_{10}} - \frac{120\sqrt{3} K_{\rm corr}^3 g_{\rm QB,c}}{\Delta_{50}\Delta_{40}\Delta_{10}} - \frac{120 K_{\rm corr}^3 g_{\rm QB,c}}{\Delta_{50}\Delta_{10}^2} - \frac{240 K_{\rm corr}^3 g_{\rm QB,c}}{\Delta_{30}\Delta_{40}\Delta_{10}} \\ &\hspace{0.5cm}- \frac{360 K_{\rm corr}^3 g_{\rm QB,c}}{\Delta_{30}\Delta_{20}\Delta_{10}} - \frac{600 K_{\rm corr}^3 g_{\rm QB,c}}{\Delta_{30}\Delta_{10}^2} - \frac{144 K_{\rm corr}^3 g_{\rm QB,c}}{\Delta_{20}\Delta_{10}^2} - \frac{144 K_{\rm corr}^3 g_{\rm QB,c}}{\Delta_{10}^3}- 12\frac{K_{\rm corr} g_{\rm QB,c}^3}{\Delta_{10}^3}
\end{split}
\end{align}
The bare coupler Hamiltonian is equivalent to a harmonic oscillator, such that the relation $\Delta_{n0} = (n-1)\Delta_{10}$ is satisfied. Therefore we can simplify the expressions for the coupling strengths
\begin{align}
J_4 &= 24\frac{g_{\rm QB,c}^4}{\Delta_{10}^3} \\
J_3 &= -6\frac{K_{\rm corr}g_{\rm QB,c}^3}{\Delta_{10}^3} \\
J_2  &= g_{\rm QB,QB} - 2\left(1 - \frac{1}{4} \frac{K_{\rm corr}}{\Delta_{10}} - \underbrace{\frac{1689+1060\sqrt{2}-82\sqrt{6}-12\sqrt{30}}{24}}_{\approx 122} \frac{K_{\rm corr}^2}{\Delta_{10}^2}\right)\frac{g_{\rm QB,c}^2}{\Delta_{10}} \\ &\hspace{0.5cm} + 40 \frac{g_{\rm QB,c}^4}{\Delta_{10}^3}\\
J_1 &= -\underbrace{\left(628+24\sqrt{3}\right)}_{\approx 670} \frac{K_{\rm corr}^3 g_{\rm QB,c}}{\Delta_{10}^3} - 12\frac{K_{\rm corr} g_{\rm QB,c}^3}{\Delta_{10}^3}
\end{align}

Putting in the expressions for $K_{\rm corr}$, $g_{\rm QB,c}$ and $\Delta_{10}$, we can write the different couplings in terms of system parameters
\begin{align}
J_4 &= 3E_{\tilde L_c}\frac{(\alpha s)^4}{\zeta_c(1-\beta_c)^{5/2}} \label{coupling1}\\
J_3  &= - E_{\tilde L_c}\frac{(\alpha s)^3 \beta_c \sqrt{\zeta_c}}{32(1-\beta_c)^3}  \label{coupling2}\\
J_2 &= E_{\tilde L_c}(\alpha s)^2 \left(1 - \frac{1}{(1-\beta_c)} + \frac{1}{2} \frac{\beta_c \zeta_c}{(1-\beta_c)^{5/2}} + c_1\frac{\beta_c^2 \zeta_c^2}{(1-\beta_c)^4} + 5 \frac{(\alpha s)^2}{\zeta_c(1-\beta_c)^{5/2}}\right)\label{coupling3},
\end{align}
with $c_1 = \frac{1 689+1060\sqrt{2}-82\sqrt{6}-12\sqrt{30}}{55296}$ and where we assumed to have identical qubits, such that $\alpha_i=\alpha_j = \alpha$, $s_i = s_j = s$. Note that all these expressions diverge for $\beta_c \longrightarrow 1$. This is since the prefactors of $H_{\rm int}$ (especially $V_{\rm QB,c}$) is in the order of $1$ in this case, such that the convergence criteria for the SW expansion is no longer satisfied. Anyways we will in the next section, that we get a really interesting effect in the regime $\beta_c < 0.6$. Note that $J_1$ and $J_3$ are two to three orders of magnitude smaller than the equal contributions hence they can be ignored (also observed in numerics).
\end{widetext}

\section{numerical evaluation of the SWT}
\label{app4}
In the previous section, we presented an analytic solution for the SWT by truncation of the cosine part of the coupler potential. To get more accurate results, it is more convenient, to include the full coupler potential and solve for the eigenfunctions numerically. We will see that corrections from higher order cosine terms play an important role for larger nonlinearities. To do so we numerically solve the Hamiltonian of the bare coupler Hamiltonian
\begin{align}
\hat H_{c} = 4 E_{\rm \tilde L_c} \xi_c^2 \frac{q_c^2}{2} + \frac{\hat \varphi}{2} + \beta_c \cos(\hat \varphi_c).
\end{align}
This Hamiltonian can be evolved in harmonic oscillator states. For the harmonic part we use the results of the previous section and the cosine part can be written down in this basis using the relation 
\begin{align}
\braket{n|{\rm e}^{ir(\hat a^{\dag} + \hat a)}|m} = {i}^{3n+m}\sqrt{\frac{n!}{m!}}{\rm e}^{-\frac{r^2}{2}}r^{n-k} L_n^{(n-m)}(r^2),
\label{app:cosformula}
\end{align}
where $L_j^{(n-m)}$ refers to the generalized Laguerre polynomial. The cosine part of the potential can now be written in the polar representation and we can write down $\hat H_c$ in the harmonic oscillator basis
\begin{align}
\hat H_c = \sum_{n,m=1}^{\infty} \braket{n|\hat H_c|m}\ket{n}\bra{m}.
\label{app:couplerseries}
\end{align}
Solving for the eigenvectors, we find the unitary transformation $\hat U_c$ that diagonalizes $\hat H_c$. With $U$ it is possible, to transform the interaction part of the Hamiltonian into the eigensystem of $\hat H_c$ (Note that the coupler parts of the interaction can easily be written down in the harmonic oscillator basis, using \eqref{app:cosformula}). This makes it easy to numerically calculate the commutators arising during the SWT by simple matrix multiplications. We truncate the series \eqref{app:couplerseries} at $n=40$ oscillator states, since higher truncation limits didn't lead to any notable changes of the results. With this we can calculate the prefactors of \eqref{app:prefactors}. We see that the results of the analytic and the numerical SWT show the same overall behavior, but the values of both are significantly different. Since the cosine part of the potential gives important contributions to the value of the energy gap, this is what we expect. For increasing $\beta_c$ the value of $\omega_c$, which denotes the gap in the analytical case decreases rapidly, pushing the calculation over the convergence limit of the SWT. By including the whole potential in the numerical case, the decrease of the gap with increasing $\beta_c$ is much slower, leading a large shift of the convergence breakdown to higher values of $\beta_c$.

\section{Discussion of the SWT}
\label{app5}
\begin{figure}
\begin{center}
\includegraphics[width=.45\textwidth]{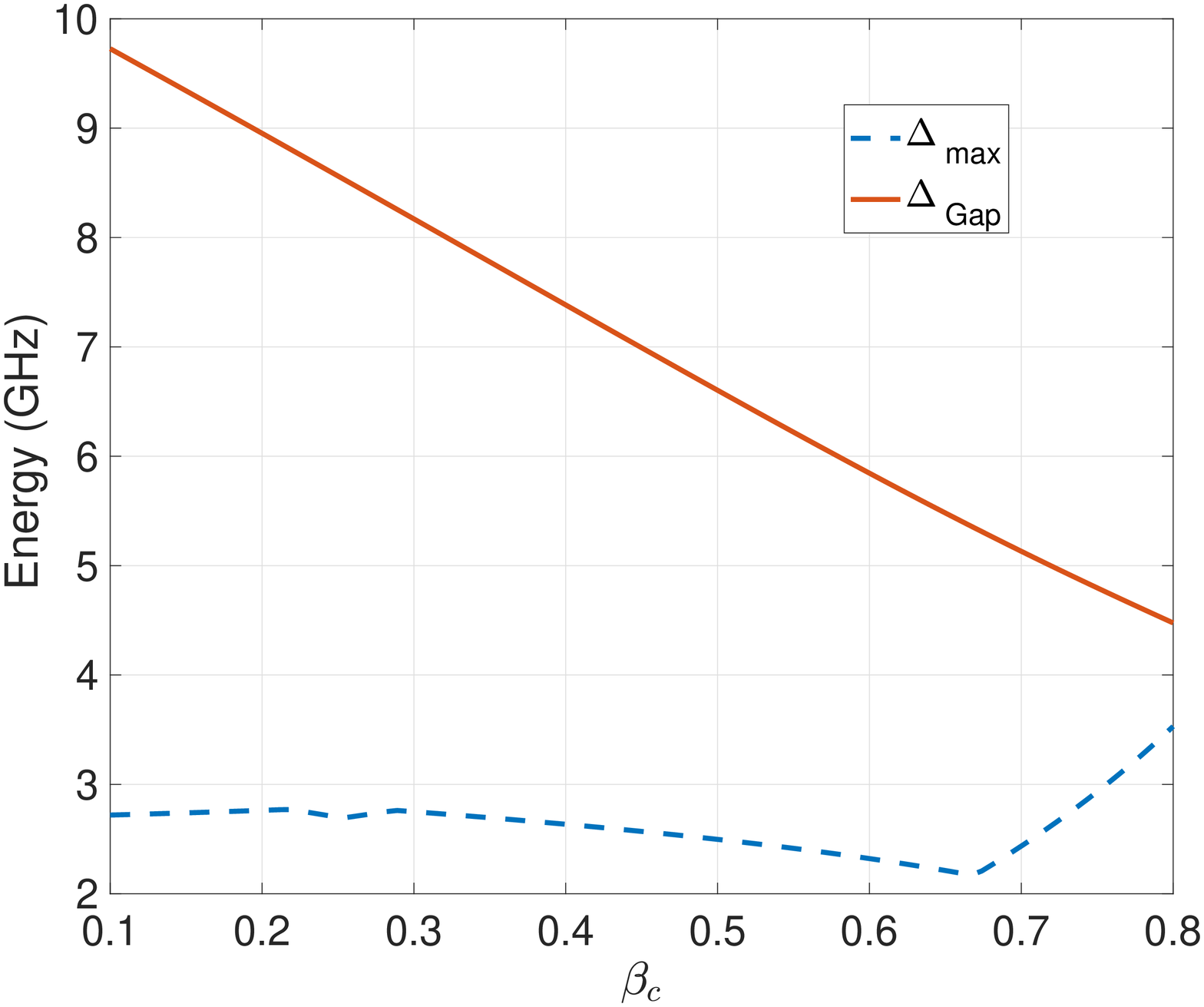}
\caption{\textit{Gap between coupler ground and coupler excited subspace $\Delta_{\rm Gap}$ and maximum spectral distance in coupler ground state subspace $\Delta_{\rm max}$ depending on the coupler nonlinearity $\beta_c$.}}
\label{app:gap}
\end{center}
\end{figure}

As mentioned in the main text, there are rather large deviations between the coupling strengths of the effective Hamiltonian obtained by the SWT and the completely numerically determined coupling strenghts. The SWT gives the right principle behavior of the coupling, meaning a change of $J_2$ from antiferromagnetic to ferromagnetic due to the indirect coupling part of the Hamiltonian and a continuous increase of $J_4$. There are two main problems why the SWT does not quantitatively model the effective system Hamiltonian. The first one is as mentioned above the way we model the qubit. As shown in Sec. \ref{app2} we model the qubit potential with two shifted harmonic potentials. The smaller the qubit nonlinearity $\beta_j$ the farer away the actual potential is from the double well. To get qualitative results we would have to include higher order corrections to the potential, but then it is no longer possible to get nice analytical results. On the other hand the more we increase the nonlinearity of the coupler $\beta_c$, the more higher orders of the SWT matter. This is because the energy distance between the coupler ground and coupler excited subspace decrease with increasing coupler nonlinearity, hence prefactors of e.g. sixth order terms increase. This is the reason for the turnover of the coupling strengths at around $\beta_c = 0.5$ (see Fig. \ref{comparison}), which we do not see in the full numerical results (see Fig. 2 of the main part). Also for to strong nonlinearities when the two subspaces start to mix, the SWT will diverge when we do not include the coupler excited subspace as well. But as soon as these subspaces start to mix, the system can no longer mimic the spectrum of the general Ising Hamiltonian including four local interactions, hence we are not interested in this regime. Here with mix we mean that the gap between the two coupler subspaces becomes comparable to the gaps of the spectral lines in the coupler ground state subspace. Then it is likely that interactions between the two subspaces happen. In Fig. \ref{app:gap} we compare the gap between the lowest state in the coupler excited and the highest state in the coupler ground subspace and the largest distance between two spectral lines in the coupler ground state subspace depending on the coupler nonlinearity $\beta_c$. For consistency reasons we chose the same qubit and coupler parameters as in Fig. 2 of the main text. One sees that for small $\beta_c$ the gap is much larger than the intersubspace energy differences, but for increasing $\beta_c$ the gap starts to decrease whereas the intersubspace energy difference increases. In between the region $0.7<\beta_c<0.8$ the two values of $\Delta_{\rm max}$ and $\Delta_{\rm gap}$ become comparable and the two subspaces are no longer well separated. This is the reason why we only show coupling strenghts up to $\beta_c = 0.7$ in Fig. 1 of the main text. 

\section{Fabrication errors}
\label{app6}

In this section we test the robustness of the our coupler against fabrication errors. For this purpose, we calculate several susceptibilities that describe the harm of fabrication errors (e.g. wrong junction parameters).

For every system parameter that arises in the four and two local interaction strength, we can define a corresponding susceptibility
\begin{align}
\chi_{4J,j}  &= \frac{1}{J_4} \sum_{\rm junctions} \left|\frac{\partial J_4}{\partial P_j}\right|_{J_4 = {\rm max}}\\
\chi_{2J,j}  &= \frac{1}{J_2} \sum_{\rm junctions} \left|\frac{\partial J_2}{\partial P_j}\right|_{J_2 = {\rm max}},
\end{align}
where the $2$ and $4$ denote the two and four local interaction strength and $P_j$ represents the system parameter that varies due to fabrication issues. Note that all our analytical results seem to only qualitatively coincide with the numerical found solution. Therefore we will calculate the susceptibilites in this article numerically. The two and four local interaction strengths can be extracted out of the spectrum and then be used to calculate the derivatives appearing in the susceptibilities. Here we assume that the optimal point is the one where the four local interaction strength is twice the two local one, so we vary the respective parameters around this optimal point.

\subsection{Error in Josephson energy}
 
\begin{figure}
\includegraphics[width=.45\textwidth]{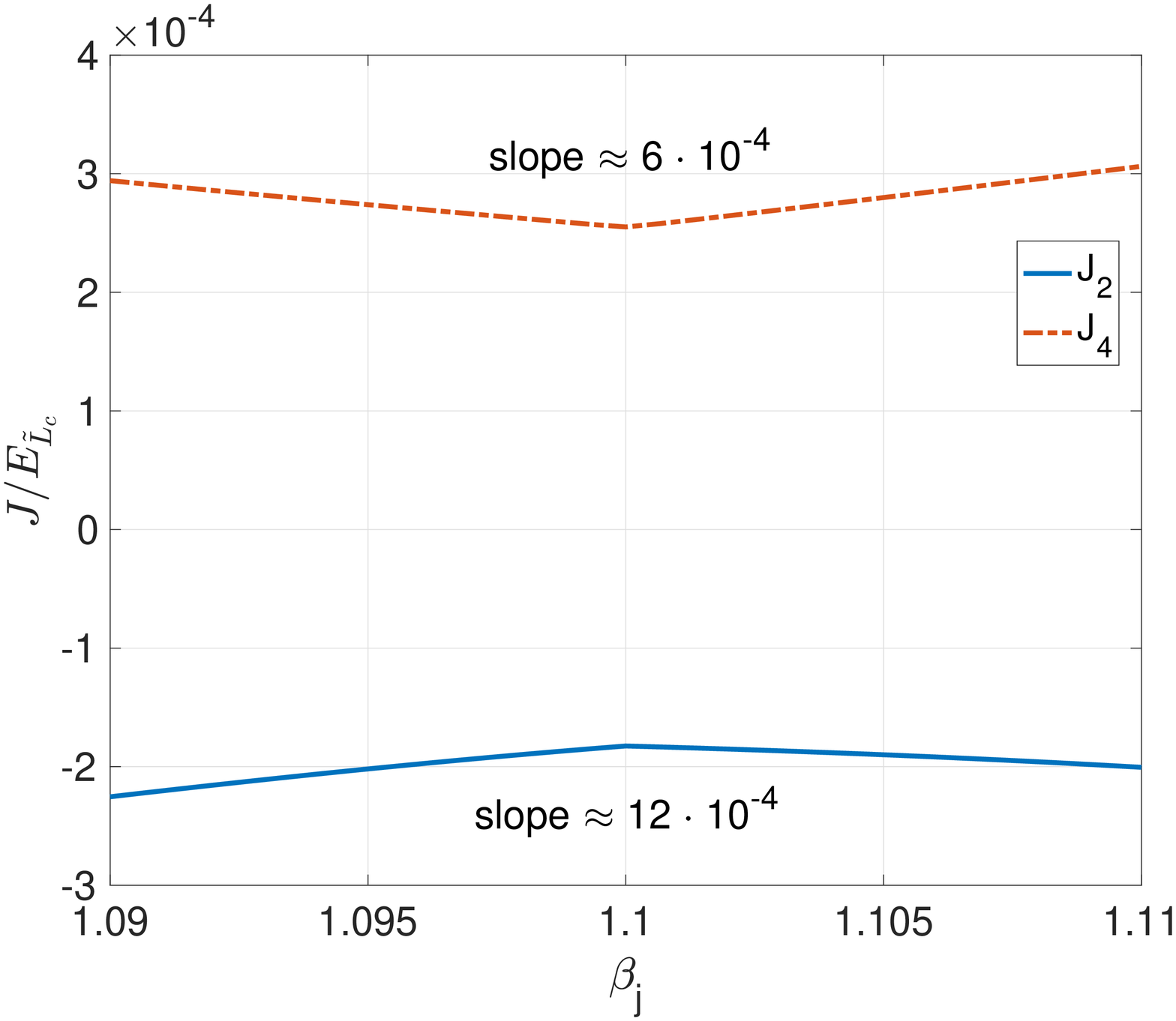}
\caption{Variation of the two and four local interactions for varying nonlinearity of the qubits. The slope can be extracted out of the plots. We choose the same parameters as in the main text $\chi_c = 0.01$, $\chi_j = 0.05$ and $\beta_c = 0.43$}
\label{susceptibility_Ejj}
\end{figure}

A typical fabrication error is an impurity in the junctions included in the system. This leads to variations of the Josephson energy. First we study a variation of the Josephson energy of the qubit junctions
\begin{align}
\chi_{4,E_{J_j}} &= \frac{4}{J_4} \left|\frac{\partial J_4}{\partial E_{J_j}}\right| \\ 
&= \frac{4}{J_4} \left|\frac{\partial J_4}{\partial \beta_j}\right|\left|\frac{\partial \beta_j}{\partial E_{J_j}}\right|\\
&= \frac{4}{J_4}\left|\frac{\partial J_4}{\partial \beta_j}\right| \frac{1}{E_{\tilde L_c}},
\end{align}
where we used the fact that only $\beta_j$ changes if we change $E_{J_j}$ and that we assume equal parameters for all four qubits (factor $4$). The derivative appearing in the expression can be calculated numerically and $E_{\tilde L_c}$ will be a normalization parameter. In Fig. \ref{susceptibility_Ejj} we show the variation of the four local and two local interactions for a small variation of $E_{J_j}$. The susceptibility for the two local interactions $\chi_{J_2,E_{J_j}}$ can be calculated analog to $\chi_{J_4,E_{J_j}}$, but we additionally have to include a factor three which arises from the fact that every qubit can interact with three others. 
Here we show the variation of the coupling strength with the nonlinearity $\beta_j$ and using Fig. \ref{susceptibility_Ejj} we can extract the derivatives $\frac{\partial J}{\partial E_{J_j}}$ we need to calculate the susceptibilites. This gives the following values for the used system parameters 
\begin{align}
E_{\rm tilde L_c} \chi_{4J,E_{J_j}} &\approx \frac{4}{J_4}  6\cdot 10^{-4} \approx 2.1 \\
E_{\rm tilde L_c} \chi_{2J,E_{J_j}} &\approx \frac{12}{J_2}  12\cdot 10^{-4} \approx 33.1.
\end{align}
We see that the two local interactions are more affected by variations of the Jospehson energies. However, $E_{\tilde L_c}$ is in the THz range for typical system parameters. This means that even for the two local interactions changing $E_{J_j}$ about 1 GHz only results in a change of the order $10^{-1}-10^{-2}$ GHz of the coupling strength. Typical fabrication errors are assumed to be much smaller than $1$ GHz, such that small variations do not crucially affect the two coupling strengths and susceptibilities are rather small. 

\begin{figure}
\includegraphics[width=.45\textwidth]{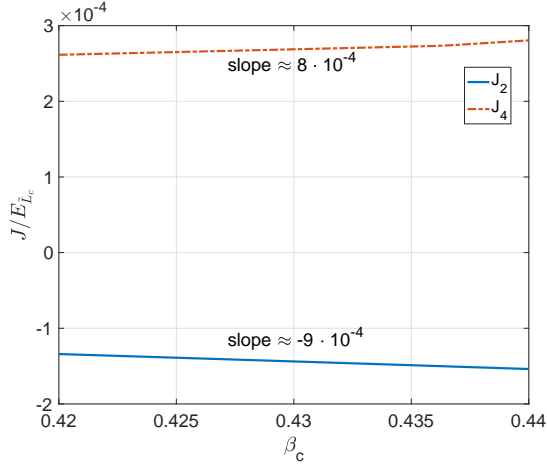}
\caption{Variation of the two and four local interactions for varying nonlinearity of the coupler. The slope can be extracted out of the plots. We choose the same parameters as in the main text $\chi_c = 0.01$, $\chi_j = 0.05$ and $\beta_c = 0.51$}
\label{susceptibility_Ejc}
\end{figure}

The same study can be done for a variation of the couplers Josephson energy. The results are shown in Fig. \ref{susceptibility_Ejc} and again we can extract the needed derivative from Fig. \ref{susceptibility_Ejc} to get an approximate value for the susceptibilities
\begin{align}
E_{\rm \tilde L_c} \chi_{4J,E_{J_c}} &\approx \frac{1}{J_4}  5\cdot 8^{-4} \approx 2.7 \\
E_{\rm \tilde L_c} \chi_{2J,E_{J_c}} &\approx \frac{1}{J_2}  4\cdot 9^{-4} \approx 6.2 .
\end{align}
As in the previous case for typical values of $E_{\tilde L_c}$ these values of the susceptibilities lead to extremely small changes of the coupling strengths when $E_{J_c}$ does not vary too much. 

\subsection{Error in inductive Energy}
\begin{figure}
\begin{center}
\includegraphics[width=.45\textwidth]{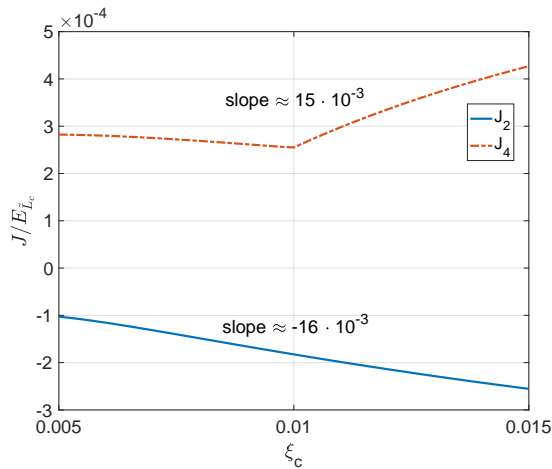}
\caption{Variation of the two and four local interactions for varying $\xi_c$. The slope can be extracted out of the plots. We choose the same parameters as in the main text $\chi_c = 0.01$, $\chi_j = 0.05$ and $\beta_c = 0.51$}
\label{susceptibility_Lc}
\end{center}
\end{figure}

Another typical fabrication error is a deviation of inductances between theoretical predicted and actual values in the experiment. In this case it is a little more difficult to calculate the corresponding susceptibilities, since more than one parameter appearing in the coupling strength depend on the impedances of the qubits $L_j$ and the coupler $L_c$, respectively. First we assume fabrication error in the coupler impedance, which means we have a change in $E_{\tilde L_c}$. The susceptibility can be written as
\begin{widetext}
\begin{align}
\chi_{4J,E_{L_c}} &= \frac{1}{J_4}\left(\left|\frac{\partial J_4}{\partial E_{\tilde L_c}}\right| \left|\frac{\partial E_{\tilde L_c}}{\partial \tilde L_c}\right|\left|{\frac{\partial}{\tilde L_c}}{\partial L_c}\right|+\left|\frac{\partial J_4}{\partial xi_c}\right| \left|\frac{\xi_c}{\partial \tilde L_c}\right|\left|{\frac{\partial}{\tilde L_c}}{\partial L_c}\right|+\left|\frac{\partial J_4}{\partial \beta_c}\right| \left|\frac{\partial \beta_c}{\partial \tilde L_c}\right|\left|{\frac{\partial}{\tilde L_c}}{\partial L_c}\right|\right)\\ 
&= \frac{1}{\tilde L_c} \left(1+\chi_c \frac{1}{J_4} \left|\frac{\partial J_4}{\partial \xi_c}\right| + \beta_c \left|\frac{\partial J_4}{\partial \beta_c}\right|\right) \\
&\Rightarrow \tilde L_c \chi_{4J,L_c} = 1+\chi_c \frac{1}{J_4} \left|\frac{\partial J_4}{\partial \xi_c}\right| + \beta_c \left|\frac{\partial J_4}{\partial \beta_c}\right|
\end{align}
\end{widetext}
Again we can plot the variation of the coupling strength around the optimal point, to numerically determine the two derivatives appearing in the expression for $\chi$. The susceptibility for the two local interactions is analog, we just have to replace $J_4$ with $J_2$. The variation with $\beta_c$ is already shown in Fig. \ref{susceptibility_Ejc} and in Fig. \ref{susceptibility_Lc} we see the variation of the coupling strengths with $\xi_c$.
For the two susceptibilities we get the approximate values
\begin{align}
\tilde L_c \chi_{4J,L_c} \approx 1.5 \\
\tilde L_c \chi_{2J,L_c} \approx 2.1.
\end{align}
However, these susceptibilities are given with respect to $\tilde L_c$. It is more convenient to look at the susceptibilites with respect to the inductive energies $E_{\tilde L_c}$ and $E_{L_j}$, respectively. We start with the first one to see how a change in $E_{\tilde L_c}$ affects the coupling strength. The corresponding susceptibility is just given by (note that we chose units such that $J_i = E_{\tilde L_c} \tilde J_i$)
\begin{align}
E_{\tilde L_c} \chi_{4J,E_{\tilde L_c}} &= 1 \\
E_{\tilde L_c} \chi_{2J,E_{\tilde L_c}} &= 4,
\end{align}
where again the factor $4$ in $J_2$ arises from the fact that four qubits interact with the coupler. A change in the inductive energy of the qubits leads to a change of $\beta_j$, such that
\begin{align}
\chi_{4J,E_{L_j}} &= \frac{4}{J_4}\left(\left|\frac{\partial J_4}{\partial \beta_j}\right|\left|\frac{\partial \beta_j}{\partial E_{L_j}}\right|\right) \\
&= \frac{4}{J_4} \frac{\beta_j}{E_{L_j}} \left|\frac{\partial J_4}{\beta_j}\right|,
\end{align}
hence we get (using Fig. \ref{susceptibility_Ejj})
\begin{align}
E_{L_j}\chi_{4J,E_{L_j}} &\approx 9 \\
E_{L_j}\chi_{2J,E_{L_j}} &\approx 36.
\end{align}
We see that these two susceptibilities are the most critical ones, since $E_{L_j}$ is one to two orders of magnitude smaller than $E_{\tilde L_c}$. Anyways, the fabrication error of inductivities is usually much smaller than the corresponding errors in the junction and we still need a huge discrepancy here to get a mentionable change of the coupling strenghts (since $E_{L_j}$ still is in the order of $10-100$ GHz). 

To summarize the susceptibility results, we have shown that only huge fabrication errors of the junctions as well as the inductances lead to significant changes of the coupling strenghts. Hence our coupler setup is assumed to be robust against fabrication errors.

\end{appendix}

\bibliography{Bibliography.bib}

\end{document}